\documentclass[a4paper,11pt]{article}
\usepackage[english]{babel}
\usepackage[ansinew]{inputenc}
\usepackage{fontenc}
\usepackage{amsfonts}
\usepackage{amsmath}
\usepackage{amsthm}
\usepackage{enumerate}
\usepackage{textcomp}
\usepackage{geometry}
\usepackage{mathrsfs}
\usepackage{natbib}
\usepackage{dsfont} % For the indicator function definition
\usepackage{graphicx}
\usepackage{multirow}

\usepackage[colorlinks=true,linkcolor=red,citecolor=blue]{hyperref}

\numberwithin{equation}{section}

\newcommand{\n}{\noindent}
\newcommand{\F}{\textbf{F}}
\newcommand{\Q}{\textbf{Q}}

\newcommand{\ind}[1]{\ensuremath{\mathds{1}_{#1}}}

\DeclareMathOperator*{\argsup}{arg\,sup\,}
\DeclareMathOperator*{\arginf}{arg\,inf\,}

\newtheorem{theorem}{Theorem}[section]
\newtheorem{definition}{Definition}[section]

\newtheorem{lemma}{Lemma}[section]
\newtheorem{proposition}{Proposition}[section]

\theoremstyle{definition}
\newtheorem{remark}{Remark}[section]
\newtheorem{example}{Example}[section]
\begin{document}

\author{Diaa AL MOHAMAD\footnote{Corresponding author. E-mail: diaa.almohamad@gmail.com} \\ \normalsize{Laboratoire de Statistique Th\'eorique et Appliqu\'ee}, \normalsize{Universit\'e Pierre et Marie Curie},\normalsize{France}}

\title{Semiparametric two-component mixture models under L-moments constraints}

\maketitle

\begin{abstract}
We propose a structure of a semiparametric two-component mixture model when one component is parametric and the other is defined through L-moments conditions. Estimation of a two-component mixture model with an unknown component is very difficult when no particular assumption is made on the structure of the unknown component. A previous work was proposed to incorporate a prior linear information concerning the distribution function of the unknown component such as moment constraints. We propose here to incorporate a prior linear information about the quantile function of the unknown component instead. This information is translated by L-moments constraints. L-moments hold better information about the tail of the distribution and are considered as good alternatives for moments especially for heavy tailed distributions since they can be defined as soon as the distribution has finite expectation. The new semiparametric mixture model is estimated using $\varphi-$divergences which permit to build feasible algorithms. Asymptotic properties of the resulting estimators are studied and proved under standard assumptions. Simulations on data generated by several mixtures models demonstrate the viability and the interest of our novel approach and the gain from using L-moment constraints in comparison to the use of moments constraints.

\n

\end{abstract}

% Theorem 4.1, Example 4.1

\section*{Introduction}
A two-component mixture model with an unknown component is defined by:
\begin{equation}
f(x) = \lambda f_1(x|\theta) + (1-\lambda) f_0(x), \qquad \text{for } x\in\mathbb{R}
\label{eqn:GeneralSemiParaMix}
\end{equation}
for $\lambda\in(0,1)$ and $\theta\in\mathbb{R}^d$ to be estimated as the density $f_0$ is unknown. This model appears in the study of gene expression data coming from microarray analysis. An application to two bovine gestation mode comparison is performed in \cite{Bordes06b}. The authors suppose that $\theta$ is known, $f_0$ is symmetric around an unknown $\mu$ and that $r=1$. \cite{Xiang} studied a more general setup by considering $\theta$ unknown and applied model (\ref{eqn:GeneralSemiParaMix}) on the Iris data by considering only the first principle component for each observed vector. Another application of model (\ref{eqn:GeneralSemiParaMix}) in genetics can be found in \cite{JunMaDiscret}. \cite{Robin} used the semiparametric model (supposing that $\theta$ is known) in multiple testing procedures in order to estimate the posterior population probabilities and the local false rate discovery. \cite{Song} studied a similar setup where $\theta$ is unknown without further assumptions on $f_0$. They applied the semiparametric model in sequential clustering algorithms as a second step after having calculated the centers of the clusters. The model of the current cluster $f_1$ is Gaussian with known location and unknown scale, whereas the distribution of the remaining clusters is represented by $f_0$ and is supposed to be unknown. Finally, Model (\ref{eqn:GeneralSemiParaMix}) can also be regarded as a contamination model, see \cite{Titterington} or \cite{FinMixModMclachlan} for further applications of mixture models. \\

Several estimation methods were proposed in the aforementioned papers without imposing any general structure on $f_0$. In the paper of \cite{Bordes06b} and its improved version by \cite{Bordes10}, the authors assume that the unknown density is symmetric around an unknown value $\mu$ and that the parametric component is fully known, i.e. $\theta$ is known. It is not possible to use this method when any of the two components of the mixture has a nonnegative support. In comparison to other existing estimation methods which do not impose any structure on the unknown component, this method has the advantage of having a solid theoretical basis. The authors prove in \cite{Bordes10} that the resulting estimators are consistent and asymptotically Gaussian. Besides, they prove (see also \cite{Bordes06b}) that the model becomes identifiable under further assumptions on the parametric component. In the paper of \cite{Song}, the authors propose two methods; the $\pi-$maximizing method and an EM-type algorithm. The $\pi-$maximizing method is based on the identifiability of model (\ref{eqn:GeneralSemiParaMix}) when $f_1$ is a scale Gaussian model. Asymptotic properties of this method were not studied and theoretical justification is only done in the Gaussian case and it is not evident how to generalize it. Their EM-type algorithm shares similarities with other exiting approaches in the literature such as \cite{Robin}, \cite{JunMaDiscret} and \cite{BordesStochEM}. These algorithms estimate at each iteration the proportion of the parametric component as an average of weights attributed to the observations. The difference between these methods is in their way of calculating the vector of weights. EM-type methods are not based on the minimization of a criterion function as in \cite{Bordes10} or \cite{Song}. Besides, their asymptotic properties are generally very difficult to establish. Finally, \cite{Xiang} propose a Hellinger-based two-step directional optimization procedure; a first step minimizes the divergence over $f_0$ and the second step minimizes over the parameters $(\lambda,\theta)$. Their method seems to give good results, but the algorithm is very complicated and no explanation on how to do the calculus is given. Properties of the iterative procedure are not studied either.\\ 

All above methods were illustrated to work in specific situations when the parametric component is fully known or is a Gaussian distribution. The paper of \cite{Xiang} shows a comparison between the methods of \cite{Bordes10}, \cite{Song} and their Hellinger-based iterative method on a two-component Gaussian data provided that the parametric component is fully known. As we add $\theta$ to the set of unknown parameters, i.e. the parametric component is not fully known, things become different. The method of \cite{Bordes10} does not perform well unless the proportion of the unknown component $1-\lambda$ is high enough. This is not surprising since this method is based on the properties of the unknown component; hence it should be well estimated. On the contrary, other methods (EM-type methods and the $\pi-$maximizing one) perform well when the proportion of the parametric component $\lambda$ is high enough.\\
It is important and of further interest that the estimation method takes into account possible unknown parameters in the parametric component $f_1$. We believe that the failure of the existing methods to treat model (\ref{eqn:GeneralSemiParaMix}) comes from the degree of difficulty of the semiparametric model itself. The use of a symmetric assumption made the estimation better in some contexts, but such assumption is still restrictive and cannot be applied on positive-supported distributions. We need to incorporate other prior information about $f_0$ in a way that we stay in between a fully parametric settings and a fully semiparametric one. \cite{DiaaAssiaMoments} have proposed a method which incorporates moment-type constraints on the unknown component. The method outperforms other semiparametric methods which do not use prior information encouraging the use of a suitable prior information. Moment-type constraints are not suitable for heavy tailed distributions. Besides, we may encounter distributions whose moments of order 2 or higher do not exist. A search for more relevent moment-type constraints is then needed.\\

We propose here to use L-moments constraints. L-moments have become classical tools alternative to central moments for the description of dispersion, skewness and kurtosis of a univariate heavy-tailed distribution. Distributions such as the Lognormal, the Pareto and the Weibull distributions are standard examples of such distributions. The use of L-moments is increasing since their introduction in \cite{Hoskings}. One of the main interests of L-moments is that they can be defined as soon as the expectation of the random variable exists. Unfortunately, the incorporation of L-moments constraints cannot be done directly in existing methods for semiparametric mixtures because the optimization will be carried over a (possibly) infinite dimensional space on the one hand, and on the other hand, existing methods use either the distribution function or the probability density function and cannot adapt an approach based on the quantile function. We thus need a new tool. Convex analysis offer away using Fenchel-Legendre duality to transform an optimization problem over an infinite dimensional space to the space of Lagrangian parameters (finite dimensional one). $\phi-$divergences offer a way by their convexity properties to use the duality of Fenchel-Legendre. \cite{AlexisGSI13} have proposed semiparametric estimation for models defined through L-moments and has shown their efficiency especially in misspecification contexts in comparison to other estimation procedures. We will combine their methodology with the work in \cite{DiaaAssiaMoments} and build a new estimation procedure which takes into account L-moments constraints over the unknown component's distribution.\\

The paper is organized as follows. Section \ref{sec:GeneralModelDef} presents L-moments and the general context of semiparametric models defined through L-moments constraints. Section \ref{sec:SPLQEstimPhiDiv} presents $\varphi-$divergences and some of their general properties. We also give briefly the estimation procedure of semiparametric models defined through L-moments constraints. Section \ref{sec:SPMixLmom} introduces semiparametric mixture models when one component is defined through L-moments conditions. We give an algorithm to estimate efficiently the parameters of the model and a plug-in estimator. Identifiability of the model and uniqueness of the estimator are also discussed. Asymptotic properties of the resulting estimator are studied and proved under standard conditions in Section \ref{sec:AsymptoticProp}. Finally, Section (\ref{sec:SimulationLmom}) is devoted to the demonstration of the method on several mixture models in univariate and multivarite contexts and a comparison with existing methods which permits to show how the prior information can improve the estimation.

%%%%%%%%%%%%%%%%%%%%%%%%%%%%%%%%%%%%%%%%%%%%%%%%%%%%%%%%%%%%%%
%
% ==========================================================================
% ==========================================================================
%
%%%%%%%%%%%%%%%%%%%%%%%%%%%%%%%%%%%%%%%%%%%%%%%%%%%%%%%%%%%%%%
\section{Semiparametric models defined through L-moments constraints}\label{sec:GeneralModelDef}
In this section, we present a definition of a semiparametric model subject to L-moments constraints. An essential part to begin with is the definition of L-moments. We will keep this part brief and one can consult \cite{AlexisThesis} Chap. 1 or \cite{Hoskings} for more details.\\
We recall two important notions; the quantile function and the quantile measure. Let $X_{1}, \ldots  X_n$ be $n$ i.i.d. copies of a random variable $X$  taking values in $\mathbb{R}$ with unknown cumulative distribution function (cdf) $\mathbb{F}$. Denote by $\mathbb{F}^{-1}(u)$ for $u \in (0,1)$ the associated quantile function of the cdf $\mathbb{F}$ defined by
\begin{equation*}
 \mathbb{F}^{-1}(u) = \inf\left\lbrace x \in \mathbb{R},\;\; s.t. \;\; \mathbb{F}(x) \geq u \right\rbrace, \;\; u \in (0,1).
\end{equation*}
We can associate to $\mathbb{F}^{-1}$ a measure ${\bf{F}}^{-1}$ on $\mathcal{B}([0,1])$ given by
\[{\bf{F}}^{-1}(B)=\int_0^1{\ind{x\in B}d\mathbb{F}^{-1}(x)} \in\mathbb{R}\cup\{-\infty,+\infty\}.\]
The integral here is a Riemann-Stieltjes one. $\F^{-1}$ is a $\sigma-$finite measure since $\mathbb{F}^{-1}$ has bounded variations on every subinterval $[a,b]$ from $(0,1)$.\\
In this section, we suppose that $\mathbb{E}|X| < \infty$ and $\int{|x|dF(x)}<\infty$. We adapt the standard notation for the cumulative distribution function (cdf) and measures, i.e. a measure $P$ has a cdf $\mathbb{F}$, a density $p$ with respect to the Lebesgue measure and a quantile measure $\F^{-1}$, and a measure $Q$ has a cdf $\mathbb{Q}$, a density $q$ with respect to the Lebesgue measure and a quantile measure $\Q$.
%%%%%%%%%%%%%%%%%%%%%%%%%%%%%%%%%%%%%%%%%%%%%%%%%%%%%%%%%%%%%%%%%%%%%%%%
%%%%%%%%%%%%%%%%%%%%%%%%%%%%%%%%%%%%%%%%%%%%%%%%%%%%%%%%%%%%%%%%%%%%%%%%
\subsection{L-moments: Definition and first properties}\label{subsec:LmomDefProper}
Let $X_{1:n} < \ldots < X_{n:n} $ be the order statistics associated to the sample $X_1,\cdots,X_n$.
\begin{definition}
The L-moment of order $r$, denoted $\lambda_r$, $r=1,2,\ldots$ is defined as a linear combination of the expectation of order statistics:
\begin{equation*}
  \lambda_r = \frac{1}{r} \sum_{k=0}^{r-1}(-1)^k \binom{r-1}{k}\mathbb{E}\left( X_{r-k:r} \right).
\end{equation*}
\end{definition}
\n If $\mathbb{F}$ is continuous, then the expectation of the $j$-th order statistic is given by
\begin{equation}
  \mathbb{E}\left[ X_{j:r} \right]  = \dfrac{r!}{(j-1)!(r-j)!} \int_{\mathbb{R}} x\mathbb{F}(x)^{j-1}\left[1-\mathbb{F}(x)\right]^{r-j}d\mathbb{F}(x).
	\label{eqn:OrderStatLaw}
\end{equation}
\n In particular, the first three L-moments are
\begin{eqnarray*}
\lambda_{1} &=& \mathbb{E}[X];\\
\lambda_{2} &=& \left( \mathbb{E}\left[X_{2:2}\right] - \mathbb{E}\left[X_{1:2}\right] \right)/2; \\
\lambda_{3} &=& \left( \mathbb{E}\left[X_{3:3}\right] - 2\mathbb{E}\left[X_{2:3}\right] + \mathbb{E}\left[X_{1:3}\right] \right)/3.
\end{eqnarray*}
\n Using formula (\ref{eqn:OrderStatLaw}), L-moments can be expressed using the quantile function $\mathbb{F}^{-1}$ (see Proposition 1.1. from \cite{AlexisThesis}) as follows:
\[\lambda_r = \int_0^1{\mathbb{F}^{-1}(u)L_{r-1}(u)du}\qquad \forall r\geq 1,\]
where $L_r$ is the shifted Legendre polynomial of order $r$ and is given by:
\[L_r(u) = \sum_{k=0}^r{(-1)^{r-k}\binom{r}{k}\binom{r+k}{k}u^k}.\]
Moreover, for $r\geq 2$:
\begin{equation}
\lambda_r = -\int_{\mathbb{R}}{K_r(t)d\mathbb{F}^{-1}(t)},
\label{eqn:LmomRepIntShiftLeg}
\end{equation}
where
\begin{equation}
K_r(t) = \int_0^t{L_{r-1}(u)du} = \sum_{k=0}^{r-1}{\frac{(-1)^{r-k}}{k+1}\binom{r}{k}\binom{r+k}{k}t^{k+1}}
\label{eqn:IntShiftLegPoly}
\end{equation}
is the integrated shifted Legendre polynomial. Notice that L-moments are polynomials in the cdf and linear in the quantile measure.

%%%%%%%%%%%%%%%%%%%%%%%%%%%%%%%%%%%%%%%%%%%%%%%%%%%%%%%%%%%%%%%%%%%%%%%%%%%%%%%%%%%%%%%%%%%%%%%%%%%%%%%
%%%%%%%%%%%%%%%%%%%%%%%%%%%%%%%%%%%%%%%%%%%%%%%%%%%%%%%%%%%%%%%%%%%%%%%%%%%%%%%%%%%%%%%%%%%%%%%%%%%%%%%

\subsection{Semiparametric Linear Quantile Models (SPLQ)}\label{subsec:SPLQDef}
SPLQ models were introduced by \cite{AlexisGSI13} (see also \cite{AlexisThesis}). The definition passes by the quantile measures instead of the distribution function. It is possible to define semiparametric models subject to L-moments constraints using the distribution function. However, their estimation would be very difficult because the constraints are not linear in the distribution function. They are instead linear in the quantile measure. This will become clearer as we go further in this subject. Denote $M^{-1}$ the set of all $\sigma-$finite measures. \\
\begin{definition} A semiparametric linear quantile model related to some quantile measure $\F_T^{-1}$ is a collection of quantile measures absolutely continuous with respect to $\F_T^{-1}$ sharing the same form of L-moments, i.e.
\[\mathcal{M} = \bigcup_{\alpha\in\mathcal{A}} \left\{\F^{-1}\ll \F_T^{-1},\; s.t.\; \int_{\mathbb{R}}{K_r(t)d\F^{-1}(t)} = m(\alpha)\right\},\]
where $m(\alpha) = (-\lambda_2,\cdots,-\lambda_{\ell})$ and $\alpha\in\mathcal{A}\subset\mathbb{R}^s$.
\end{definition}
\begin{example}[\cite{AlexisThesis}]
\label{Example:Weibull}
Consider the model which is the family of all the distributions of a r.v. $X$ whose second, third and fourth L-moments satisfy:
\begin{eqnarray*}
\lambda_{2} &=& \sigma\left( 1-2^{-1/\nu} \right)\Gamma\left( 1 + \dfrac{1}{\nu} \right);\\
\lambda_{3} &=& \lambda_{2}\left[ 3 - 2\dfrac{1-3^{1/\nu}}{1-2^{-1/\nu}} \right];\\
\lambda_{4} &=& \lambda_{2}\left[ 6 + \dfrac{5(1-4^{-1/\nu})-10(1-3^{-1/\nu})}{1-2^{-1/\nu}} \right],
\end{eqnarray*}
for $\sigma > 0$, $\nu > 0$. These distributions share their first L-moments of order 2, 3 and 4 with those of a Weibull distribution with scale and shape parameters $\sigma$, $\nu$.
\end{example}
%\textcolor{blue}{For constrained optimization problems with linear constraints, convex analysis offers a tool using the Fenchel duality to solve such optimization problems as long as the objective function is convex. In the previous chapter, we exploited this tool to transform the optimization of the $\varphi-$divergence under moment-type constraints in possibly infinite dimensional space into an optimization problem over the finite dimensional space $\mathbb{R}^{\ell-1}$. L-moments constraints, however, are \emph{not} linear functionals of the cdf. They are still linear functionals of the quantile measures. Based upon this idea, \cite{AlexisThesis} built a mathematical procedure to treat efficiently semiparametric models constrained to L-moments conditions. }
%
%
In SPLQ models, the objective is to estimate the value of $\alpha^*$ for which the true quantile measure $\F^{-1}_T$ of the data belongs to $\mathcal{M}_{\alpha^*}$ on the basis of a sample $X_1,\cdots,X_n$. The estimation procedure is generally done by either solving the set of equations defining the constraints or by minimizing a suitable distance-like function between the set $\mathcal{M}$ and some estimator of $\F^{-1}_T$ based on an observed sample. In other words, we search for the "projection" of $\F^{-1}_T$ on $\mathcal{M}$.

%%%%%%%%%%%%%%%%%%%%%%%%%%%%%%%%%%%%%%%%%%%%%%%%%%%
% ===============================================
%%%%%%%%%%%%%%%%%%%%%%%%%%%%%%%%%%%%%%%%%%%%%%%%%%%%
\section{Estimation of SPLQ models using \texorpdfstring{$\varphi$}{phi}-divergences}\label{sec:SPLQEstimPhiDiv}
\subsection{Definitions and properties}\label{subsec:DefPropPhiDiverg}
$\varphi$-divergences were introduced independently by \cite{Csiszar1963} (as "$f$-divergences") and \cite{AliSilvey}. Let $P$ and $Q$ be two $\sigma-$finite measures defined on $(\mathbb{R}^{r}, \mathscr{B}(\mathbb{R}^{r}))$ such that $Q$ is absolutely continuous (a.c.) with respect to (w.r.t.) $P$. Let $\varphi : \mathbb{R} \mapsto [0, +\infty]$ be a proper convex function with $\varphi(1) = 0$ and such that its domain $\textrm{dom}\varphi = \left\lbrace   x \in \mathbb{R} \;\; \textrm{such that} \;\; \varphi(x) < \infty \right\rbrace := (a_{\varphi},b_{\varphi})$ with $a_{\varphi} < 1 < b_{\varphi}$. The $\varphi$-divergence between $Q$ and $P$ is defined by:

\begin{equation*}
D_{\varphi}(Q,P) = \int_{\mathbb{R}^r}{ \varphi\left( \dfrac{dQ}{dP}(x) \right)dP(x)},
\end{equation*}

\n
where $\dfrac{dQ}{dP}$ is the Radon-Nikodym derivative. When $Q$ is not a.c.w.r.t. $P$, we set $D_{\varphi}(Q,P) = + \infty$. When, $P = Q$ then $D_{\varphi}(Q,P) = 0$. Furthermore, if the function $x \mapsto \varphi(x)$ is strictly convex on a neighborhood of $x=1$, then

\begin{equation}
\label{fondamental property of divergence}
D_{\varphi}(Q,P) = 0 \;\; \textrm{ if and only if} \;\; P = Q.
\end{equation}
In our work, $P$ and $Q$ will be quantile measures. Several standard statistical divergences can be expressed as $\varphi-$divergences; the Hellinger, the Pearson's and the Neymann's $\chi^2$, and the (modified) Kullback-Leibler. They all belong to the class of Cressie-Read \cite{CressieRead1984} (also known as "power divergences") defined by:
\begin{equation}
\varphi_{\gamma}(x) := \dfrac{x^{\gamma}-\gamma x + \gamma -1}{\gamma(\gamma -1)},
\label{eqn:CressieReadPhi}
\end{equation}
for $\gamma=\frac{1}{2},2,-2,0,1$ respectively\footnote{For $\gamma\in\{0,1\}$, the limit is calculated since it is not well-defined. We denote $\varphi_0(t)=-\log(t)+t-1$ and $\varphi_1(t)=t\log(t)-t+1$.}.
More details and properties can be found in \cite{LieseVajda} or \cite{Pardo}. \\
Estimators based on $\varphi-$divergences were developed in the parametric (see \cite{Beran},\cite{LindsayRAF},\cite{ParkBasu},\cite{BroniaKeziou09}) and the semiparametric setups (see \cite{BroniaKeziou12} and \cite{AlexisGSI13}). In all these methods, the $\varphi-$divergence is calculated between a model $Q$ and a true distribution $P_T$. We search our estimators by making the model approaches\footnote{More accurately, we project the true distribution on the model.} the true distribution $P_T$. In this paper, we provide an original method where the minimization is done over both arguments of the divergence in a way that the two arguments approach one another for the sake of finding a suitable estimator\footnote{There still exists some work in computer vision using $\varphi-$divergences where the minimization is done over both arguments of the divergence, see \cite{Mireille}.  The work concerns a parametric setup in discrete models.}. In completly nonparametric setup, we may mention the work of \cite{KarunamuniWu} on two component mixture models when both components are unknown. The authors use the Hellinger divergence, and assume that we have in hand a sample from each component and a sample drawn from the whole mixture. For regression of nonparametric mixture models using the Hellinger divergence, see the paper of \cite{TangRegression}.\\
The following definitions concern the notion of $\varphi$-projection of $\sigma-$finite measures over a set of $\sigma-$finite measures.

\begin{definition}
\label{def:phiDistance}
Let $\mathcal{M}$ be some subset of $M^{-1}$, the space of $\sigma-$finite measures. The $\varphi$-divergence between the set $\mathcal{M}$ and some $\sigma-$finite measure $P$, noted as $D_{\varphi}(\mathcal{M},P)$, is given by
\begin{equation}
  D_{\varphi}\left(\mathcal{M},P\right)  := \inf_{Q \in \mathcal{M}}D_{\varphi}(Q,P).
	\label{eqn:PhiDistanceEleSet}
\end{equation}
Furthermore, we define the $\varphi-$divergence between two subsets of $M^{-1}$, say $\mathcal{M}$ and $\mathcal{N}$ by:
\begin{equation*}
  D_{\varphi}\left(\mathcal{M},\mathcal{N}\right)  := \inf_{Q \in \mathcal{M}}\inf_{P\in\mathcal{N}}D_{\varphi}(Q,P).
\end{equation*}
\end{definition}

\begin{definition}
Assume that $D_{\varphi}(\mathcal{M},P)$ is finite. A measure $Q^* \in \mathcal{M}$ such that
\begin{equation*}
D_{\varphi}(Q^*,P) \leq D_{\varphi}(Q,P), \;\; \textrm{for all}\;\; Q \in \mathcal{M}
\end{equation*}
is called a $\varphi$-projection of $P$ onto $\mathcal{M}$. This projection may not exist, or may not be defined uniquely.
\end{definition}

The essential tool we need from a $\varphi-$divergence is its characterization of the projection of some $\sigma-$finite measure $P$ onto a set $\mathcal{M}$ of $\sigma-$finite measures. Such characterization will permit to transform the search of a projection in an infinite dimensional space to the search of a vector $\xi$ in $\mathbb{R}^{\ell-1}$.
%%%%%%%%%%%%%%%%%%%%%%%%%%%%%%%%%%%%%%%%%%%%%%%%%%%%%%%%%%%%%%%%%%%%%%%%%%%%
\subsection{Estimation of SPLQ models using \texorpdfstring{$\varphi-$}{phi}divergences and the duality technique}\label{subsec:SPLQDuality}
Estimation of SPLQ models using $\varphi-$divergences is summarized by the following optimization problem:
\begin{eqnarray}
\alpha^* & = & \arginf_{\alpha\in\mathcal{A}} D_{\varphi}\left(\mathcal{M}_{\alpha},\F_T^{-1}\right) \nonumber \\
 & = & \arginf_{\alpha\in\mathcal{A}} \inf_{\F^{-1}\in\mathcal{M}_{\alpha}}D_{\varphi}\left(\F^{-1},\F_T^{-1}\right).
\label{eqn:EstimSPLQ}
\end{eqnarray}
In other words, we are looking for the projection of $\F_T^{-1}$ on the set $\mathcal{M}$ by minimizing the $\varphi-$divergence between them. Of course, if $\F_T^{-1}\in\mathcal{M}_{\alpha^*}$ for some $\alpha^*\in\mathcal{A}$, then $D_{\varphi}\left(\cup_{\alpha}\mathcal{M}_{\alpha},\F_T^{-1}\right)=0$. Otherwise, $\alpha^*$ corresponds to the parameter of the closest set $\mathcal{M}_{\alpha}$ from the $\varphi-$divergence point of view to the quantile measure $\F_T^{-1}$.\\
The estimation procedure (\ref{eqn:EstimSPLQ}) is not feasible because it concerns the minimization over a subset of possibly infinite dimensional space. The duality technique permits to transform the calculus of the projection from an optimization problem over a possibly infinite dimensional space into an optimization problem over $\mathbb{R}^{\ell-1}$, where $\ell-1$ is the number of constraints defining the set $\mathcal{M}_{\alpha}$. We recall briefly this technique by applying it directly in the context of quantile measures. Corollary 1.1 from \cite{AlexisThesis} states the following. If there exists some $\F^{-1}$ in $\mathcal{M}_{\alpha}$ such that $a_{\varphi}<d\F^{-1}/d\F_T^{-1}<b_{\varphi}$ $\F_T^{-1}$-a.s. where dom$\varphi = (a_{\varphi},b_{\varphi})$ then,
\begin{equation}
\inf_{\F^{-1}\in\mathcal{M}_{\alpha}}\int_{0}^1{\varphi\left(\frac{d\F^{-1}}{d\F_T^{-1}}\right)d\F_T^{-1}} = \sup_{\xi\in\mathbb{R}^{\ell-1}} \xi^t m(\alpha) - \int_0^1{\psi\left(\xi^tK(u)\right)d\F_T^{-1}(u)}.
\label{eqn:DualityLmom}
\end{equation}
This formula permits to build a plug-in estimate for $\alpha$ by considering a sample $X_1,\cdots,X_n$, see Remark 1.15 in \cite{AlexisThesis}.
\begin{equation}
\hat{\alpha} = \arginf_{\alpha} \sup_{\xi\in\mathbb{R}^{\ell-1}} \xi^t m(\alpha) - \sum_{i=1}^{n-1}{\psi\left(\xi^tK\left(\frac{i}{n}\right)\right)\left(X_{i+1:n} - X_{i:n}\right)}.
\label{eqn:DualityLmomEmpirical}
\end{equation}
Asymptotic properties of this estimator were studied in \cite{AlexisGSI13} (see also \cite{AlexisThesis} Theorems 1.2 and 1.3). This plug-in estimate is very interesting in its own, because it does not need any numerical integration. Besides, if we take $\varphi$ to be the $\chi^2$ generator, i.e. $\varphi(t)=(t-1)^2/2$ whose convex conjugate is $\psi(t)=t^2/2+t$, the optimization over $\xi$ can be solved directly, see Example 1.12 in \cite{AlexisThesis}. We will get back to this interesting case study later on.\\
Now that all notions and analytical tools are presented, we proceed to the objective of this paper; semiparametric mixtures models. The following section defines such models and presents our proposed method to estimate them using $\varphi-$divergences. It follows then the plug-in estimates and their asymptotic properties.

%%%%%%%%%%%%%%%%%%%%%%%%%%%%%%%%%%%%%%%%%%%%%%%%%%%%%%%%%%%%%%
%
% ==========================================================================
% ==========================================================================
%
%%%%%%%%%%%%%%%%%%%%%%%%%%%%%%%%%%%%%%%%%%%%%%%%%%%%%%%%%%%%%%
\section{Semiparametric two-component mixture models when one component is defined through L-moments constraints}\label{sec:SPMixLmom}

\subsection{Definition and identifiability}
\begin{definition}
\label{def:SemiParaModelLmom}
Let $X$ be a random variable taking values in $\mathbb{R}$ distributed from a probability measure $P$ whose cdf is $\mathbb{F}$. We say that $P(.|\phi)$ with $\phi=(\lambda,\theta,\alpha)$ is a two-component semiparametric mixture model subject to L-moments constraints if it can be written as follows:
\begin{eqnarray}
P(.| \phi) & = &  \lambda P_1(.|\theta) + (1-\lambda) P_0 \quad \text{s.t. } \nonumber\\
\F_0^{-1}\in\mathcal{M}_{\alpha} & = & \left\{\Q^{-1} \in M^{-1}, \Q^{-1}\ll\F_0^{-1} \text{ s.t. } \int_{0}^1{K(u)d\Q^{-1}}=m(\alpha)\right\}
\label{eqn:SetMalphaLmom}
\end{eqnarray}
for $\lambda\in(0,1)$ the proportion of the parametric component, $\theta\in\Theta\subset\mathbb{R}^{d}$ a set of parameters defining the parametric component, $\alpha\in\mathcal{A}\subset\mathbb{R}^{s}$ is the constraints parameter, $K=(K_2,...,K_{\ell})$ is defined through formula (\ref{eqn:IntShiftLegPoly}) and finally $m(\alpha)=(m_2(\alpha),\cdots,m_{\ell}(\alpha))$ is a vector-valued function determining the values of the L-moments.
\end{definition}
Notice that $m(\alpha)$ is must contain the negative values of the L-moments by equation (\ref{eqn:LmomRepIntShiftLeg}), i.e $m_r(\alpha)=-\lambda_r$. In this definition, it may appear that we have mixed quantiles with probabilities. This is however necessary in order to show the structure of the mixture model which generates the data. This structure is uniquely defined through the distribution function and does not have a "proper" writing using the quantile measure. In general, there is no formula which gives the quantile of a mixture model, and in practice, statisticians use approximations to calculate the quantile of a mixture model. Thus, working with the quantiles will make us lose the linearity property relating the two components with the mixture's distribution. In the semiparametric two-component mixture model (\cite{DiaaAssiaMoments}), this linearity played an essential role in the estimation procedure and simplified the calculus of the estimator on several levels. We will get back to this idea later on, and a "partial" solution will be proposed in order to get back to work with the cdf instead of its quantile.\\
It is important to recall that the use of quantiles in the definition of semiparametric models subject to L-moments constraints stems from the fact that the constraints are linear functionals in the quantiles. Thus, an estimation procedure which employs the quantiles instead of the distribution function can be solved using the Fenchel-Legendre duality in a similar way to paragraph \ref{subsec:SPLQDuality}.

%%%%%%%%%%%%%%%%%%%%%%%%%%%%%%%%%%%%%%%%%%%%%%%%%%%%%%%
%\subsection{Identifiability}
The identifiability of the model was not questioned in the context of SPLQ models because it suffices that the sets $\mathcal{M}_{\alpha}$ are disjoint (the function $m(\alpha)$ is one-to-one). However, in the context of this semiparametric mixture model, identifiability cannot be achieved only by supposing that the sets $\mathcal{M}_{\alpha}$ are disjoint. \\
\begin{definition}
We say that the two-component semiparametric mixture model subject to L-moments constraints is identifiable if it verifies the following assertion. If
\begin{equation}
\lambda P_1(.|\theta) + (1-\lambda)P_0 = \tilde{\lambda} P_1(.|\tilde{\theta}) + (1-\tilde{\lambda}) \tilde{P}_0,\quad \text{with } \F_0^{-1}\in\mathcal{M}_{\alpha}, \tilde{\F}_0^{-1}\in\mathcal{M}_{\tilde{\alpha}}, 
\label{eqn:IdenitifiabilityDefEqLmom}
\end{equation}
then $\lambda = \tilde{\lambda},\theta = \tilde{\theta}$ and $P_0=\tilde{P}_0$ (and hence $\alpha=\tilde{\alpha}$).
\end{definition}
This is the same identifiability concept considered by \cite{Bordes06b} except that, in our definition the unknown component's quantile belongs to the set $\mathcal{M}_{\alpha}$.\\
\begin{proposition}
\label{prop:identifiabilityMixtureLmom}
For a given mixture distribution $P_T = P(.| \phi^*)$ whose cdf is $\mathbb{F}_T$, suppose that the system of equations:
\begin{equation}
\int_{0}^1{K(u)d\left(\frac{1}{1-\lambda}\F_T - \frac{\lambda}{1-\lambda}\F_1(.|\theta)\right)^{-1}(u)} = m(\alpha)
\label{eqn:SysLmomEq}
\end{equation}
has a unique solution $(\lambda^*,\theta^*,\alpha^*)$. Then, equation (\ref{eqn:IdenitifiabilityDefEqLmom}) has a unique solution, i.e. $\lambda = \tilde{\lambda},\theta = \tilde{\theta}$ and $P_0=\tilde{P}_0$, and the semiparametric mixture model $P_T = P(.| \phi^*)$ is identifiable.
\end{proposition}
The proof is differed to Appendix \ref{AppendSemiPara:PropIdenitifiabilityLmom}.
%Example
\begin{example}[Two-component exponential mixture]
We propose to look at an exponential mixture defined by:
\[f(x|\lambda^*,a_1^*) = \lambda^* a_1^* e^{-a_1^* x} + (1-\lambda^*) a_0^* e^{-a_0^* x}\]
where $a_1^*=1.5, a_0^*=0.5$ and $\lambda^*\in\{0.3,0.5,0.7,0.85\}$. This is considered to be the distribution generating the observed data. Suppose that the second component $f_0^*(x)=a_0^* e^{-a_0^* x}$ is unknown during the estimation. Furthermore, suppose that we hold an information about $f_0^*$ that its quantile ${\F_0^*}^{-1}$ belongs to the following class of functions:
\[\mathcal{M} = \left\{\F^{-1}\ll{\F_0^*}^{-1},\quad \int_0^1{u(1-u)d\F^{-1}(u)} = \frac{1}{2a_0^*}\right\}.\]
%\int_{\mathbb{R}_+}{(2\mathbb{F}_0(x)-1)f_0(x)dx}
This set contains all probability distributions whose second L-moment has the value $\frac{1}{2a_0^*}$. We would like to check the identifiability of the semiparametric mixture model subject to the second L-moment constraint of the exponential distribution $\mathcal{E}(a_0^*)$. The system of equations (\ref{eqn:SysLmomEq}) is given by:
\[\int_0^1{u(1-u)d\left(\frac{1}{1-\lambda}\F_T - \frac{\lambda}{1-\lambda}\F_1(.|a_1)\right)^{-1}(u)} = \frac{1}{2a_0^*}.\]
In order to calculate the left hand side, we use the alternative definition of the second L-moment $\lambda_2 = \left(\mathbb{E}[X_{2:2}] - \mathbb{E}[X_{1:2}]\right)/2$ and exploit formula (\ref{eqn:OrderStatLaw}). We have
\begin{multline*} \int_0^1{u(1-u)d\left(\frac{1}{1-\lambda}\F_T - \frac{\lambda}{1-\lambda}\F_1(.|a_1)\right)^{-1}(u)} = \\ 
\int_{\mathbb{R}_+}{x\left[2\left(\frac{1}{1-\lambda}\F_T(x) - \frac{\lambda}{1-\lambda}\F_1(x|a_1)\right)-1\right]\left(\frac{1}{1-\lambda}p_T(x) - \frac{\lambda}{1-\lambda}p_1(x|a_1)\right)dx} 
\end{multline*}
A direct calculus of the right hand side shows:
\begin{multline*}
\int_{\mathbb{R}_+}{x\left[2\left(\frac{1}{1-\lambda}\F_T(x) - \frac{\lambda}{1-\lambda}\F_1(x|a_1)\right)-1\right]\left(\frac{1}{1-\lambda}p_T(x) - \frac{\lambda}{1-\lambda}p_1(x|a_1)\right)dx}  = \\ \frac{2C_1-(\lambda+1)C_2}{(1-\lambda)^2} + \frac{\lambda^2-2\lambda}{2a_1(1-\lambda)^2} + \frac{2\lambda^*\lambda}{(1-\lambda)^2(a_1+a_1^*)} + \frac{2\lambda(1-\lambda^*)}{(1-\lambda)^2(a_1+a_0^*)}
\end{multline*}
where
\begin{eqnarray*}
C2 & = & \frac{\lambda^*}{a_1^*} + \frac{1-\lambda^*}{a_0^*} \\
C1 & = & \frac{\lambda^*}{a_1^*} + \frac{1-\lambda^*}{a_0^*} - \frac{{\lambda^*}^2}{4a_1^*} - \frac{(1-\lambda^*)^2}{4a_0^*} - \frac{\lambda^*(1-\lambda^*)}{a_1^*+a_0^*}.
\end{eqnarray*}
In figure (\ref{fig:2ndLmomConstr}), we show the set of solutions of the following equation:
\begin{equation}
\frac{2C_1-(\lambda+1)C_2}{(1-\lambda)^2} + \frac{\lambda^2-2\lambda}{2a_1(1-\lambda)^2} + \frac{2\lambda^*\lambda}{(1-\lambda)^2(a_1+a_1^*)} + \frac{2\lambda(1-\lambda^*)}{(1-\lambda)^2(a_1+a_0^*)} = \frac{1}{2a_0^*},
\label{eqn:LmomentExpoEquation}
\end{equation}
for several values of $\lambda^*$ in the figure to the left. The figure to the right shows the intersection between the set of solutions and the set $\Phi^+=\{(\lambda,a), \text{ s.t. }\frac{1}{1-\lambda}\F_T(x) - \frac{\lambda}{1-\lambda}\F_1(x|a_1) \text{ is a cdf}\}$. It is clear that the nonlinear system of equations (\ref{eqn:SysLmomEq}) has an infinite number of solutions. In order to reduce the number of solutions into one, we need to consider another L-moment constraint. We do not pursue this here because the calculus is already complicated even in this simple model.\\ 
Note that the set of solutions is shrinking as the proportion of the unknown component $f_0$ becomes smaller (the value of $\lambda^*$ increases). This gives rise to a difficult and an important question; what happens if we have a number of constraints inferior to the number of parameters. This question is not pursued here.
\begin{figure}[ht]
\centering
\includegraphics[scale=0.45]{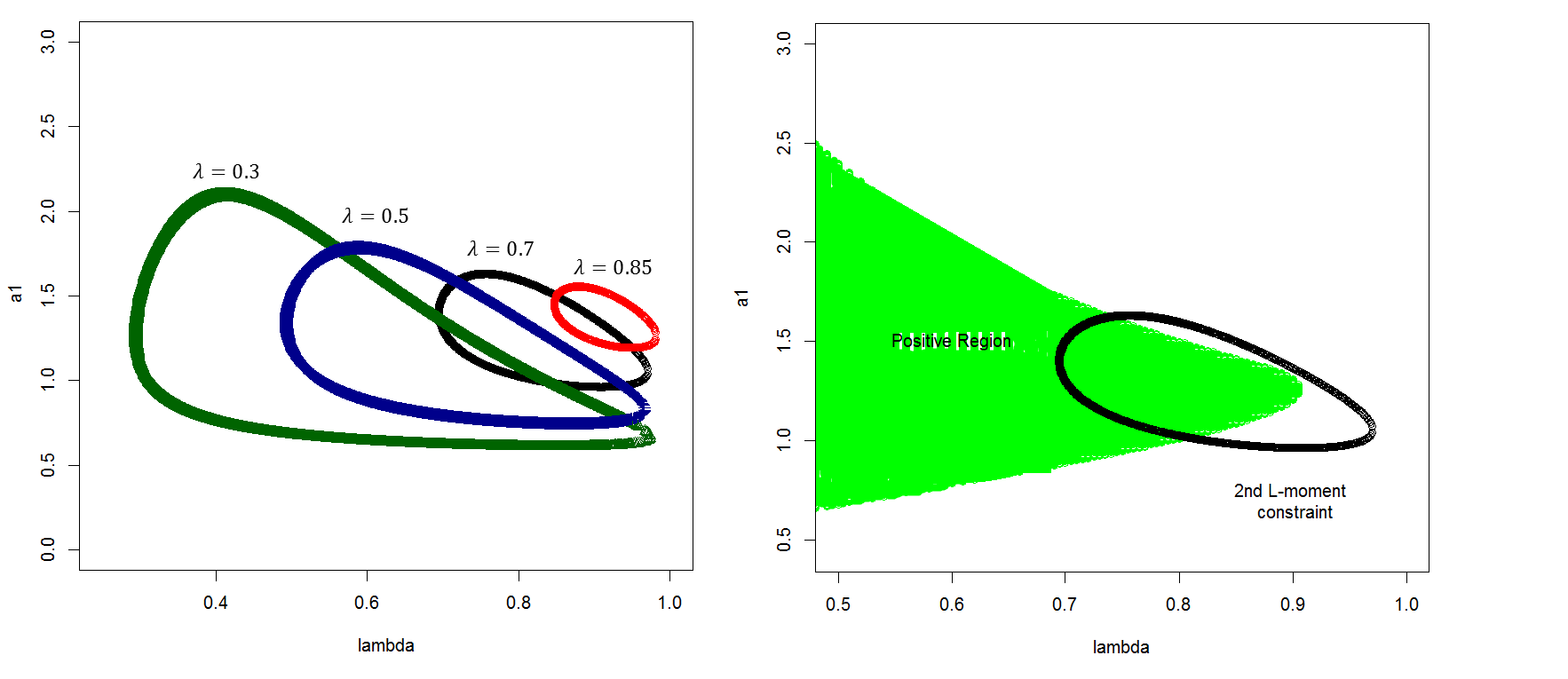}
\caption{The set of solutions under a constraint over the second L-moment. Each closed trajectory corresponds to a value of the proportion of the parametric part indicated above of it. The figure to the left represents the whole set of solutions of the equation (\ref{eqn:LmomentExpoEquation}) for different values of the true proportion $\lambda^*$. The figure to the right represents the intersection between the set of solutions of equation (\ref{eqn:LmomentExpoEquation}) for $\lambda^*=0.7$ with the set $\Phi^+$.}
\label{fig:2ndLmomConstr}
\end{figure}

\end{example}
%%%%%%%%%%%%%%%%%%%%%%%%%%%%%%%%%%%%%%%%%%%%%%%%%%%%%%%
\subsection{An algorithm for the estimation of the semiparametric mixture model}
In the context of our semiparametric mixture model, we want to estimate the parameters $(\lambda,\theta,\alpha)$ on the basis of two pieces of information; an i.i.d. sample $X_1,\cdots,X_n$ drawn from $P_T$ and the fact that ${\F_0^*}^{-1}$ belongs to the set $\mathcal{M}$. For SPLQ models, we have seen that using $\varphi-$divergences, we were able to construct an estimation procedure by minimizing some distance between the set of constraints and the distribution generating the data. The resulting estimation procedure is an optimizing problem over an infinite dimensional space. We exploited the linearity of the constraints and transformed the estimation procedure into a feasible optimization problem over $\mathbb{R}^{\ell-1}$ using the Fenchel-Legendre duality. \\
In order to use the Fenchel-Legendre duality, the constraints need to apply over the whole mixture. In our semiparametric mixture model, the constraints apply over the quantile of only one component; $\F_0^{-1}$. We thus propose to define another "model" based on $\F_0^{-1}$. We have:
\begin{equation}
{\mathbb{F}_0^*}^{-1} = \left(\frac{1}{1-\lambda^*}\mathbb{F}_T(.|\phi^*) - \frac{\lambda^*}{1-\lambda^*}\mathbb{F}_1(|\theta^*)\right)^{-1}.
\label{eqn:TrueF0model}
\end{equation}
Define the set $\mathcal{N}^{-1}$ by:
\[\mathcal{N}^{-1} = \left\{\Q^{-1} \in M^{-1}\; : \; \exists (\lambda,\theta)\in(0,1)\times\Theta \text{ s.t. } \mathbb{Q}^{-1} = \left(\frac{1}{1-\lambda}\mathbb{F}_T - \frac{\lambda}{1-\lambda}\mathbb{F}_1(.|\theta)\right)^{-1}\right\}.\]
Notice that not all the couples $(\lambda,\theta)$ in $(0,1)\times\Theta$ are accepted, because function $\frac{1}{1-\lambda}\mathbb{F}_T - \frac{\lambda}{1-\lambda}\mathbb{F}_1(|\theta)$ may not be a cdf for for these couples. Define the set of effective parameters $\Phi^+$ by:
\begin{equation}
\Phi^+=\left\{(\lambda,\theta)\in(0,1)\times\Theta : \frac{1}{1-\lambda}\mathbb{F}_T - \frac{\lambda}{1-\lambda}\mathbb{F}_1(.|\theta) \text{ is a cdf}\right\}.
\end{equation}
Now, the set $\mathcal{N}^{-1}$ can be characterized using $\Phi^+$ by:
\[\mathcal{N}^{-1} = \left\{\left(\frac{1}{1-\lambda}\F_T - \frac{\lambda}{1-\lambda}\F_1(.|\theta)\right)^{-1}, \text{ for } (\lambda,\theta)\in\Phi^+\right\}.\]
%A simple check can still be done. It is simple to see that if the associated "density function" $\frac{1}{1-\lambda}p_T - \frac{\lambda}{1-\lambda}p_1(|\theta)$ is positive (hence, a pdf), then function $\frac{1}{1-\lambda}\mathbb{F}_T - \frac{\lambda}{1-\lambda}\mathbb{F}_1(|\theta)$ is a cdf. We have seen in the case of moment-type constraints, that the effective set of parameters $\Phi^+$ is governed by the matrix $J_H$ being definite negative which can be fulfilled if $\frac{1}{1-\lambda}p_T - \frac{\lambda}{1-\lambda}p_1(|\theta)$ is positive. This will not create any difficulty or loss of information in the final step of our procedure as we will explain later. It is the inverse. We will be able to use the whole set of parameters $\Phi$ instead of only the set where the matrix $J_H$ is definite negative.\\
%For the time being, we are obliged in order to give a proper introduction of our estimation procedure to work with couples $(\lambda,\alpha)$ which makes $\frac{1}{1-\lambda}\mathbb{F}_T - \frac{\lambda}{1-\lambda}\mathbb{F}_1(|\theta)$ a cdf.\\
Notice that ${\F_0^*}^{-1}$ is a member of $\mathcal{N}^{-1}$ for $(\lambda,\theta)=(\lambda^*,\theta^*)$. On the other hand, and by definition of the semiparametric mixture model, ${\F_0^*}^{-1} \in\mathcal{M}_{\alpha^*}$. We may write:
\begin{equation}
{\F_0^*}^{-1}\in \mathcal{N}^{-1} \bigcap \cup_{\alpha}\mathcal{M}_{\alpha}.
\label{eqn:F0InIntersect}
\end{equation}
If we suppose that the intersection (which is not void) contains only one element which will be ${\F_0^*}^{-1}$, then it becomes reasonable to consider an estimation procedure by calculating a "distance" between the two sets $\cup_{\alpha}\mathcal{M}_{\alpha}$ and $\mathcal{N}^{-1}$. Using definition \ref{def:phiDistance}, we may write:
\begin{eqnarray}
D_{\varphi}\left(\cup_{\alpha}\mathcal{M}_{\alpha}, \mathcal{N}^{-1}\right) & = & \inf_{\Q^{-1}\in\mathcal{N}^{-1}}\inf_{\F_0^{-1}\in\cup_{\alpha}\mathcal{M}_{\alpha}} D_{\varphi}\left(\F_0^{-1},\Q^{-1}\right) \nonumber\\
 & = & \inf_{(\lambda,\theta)\in\Phi^+,\alpha\in\mathcal{A}}\inf_{\F_0^{-1}\in\mathcal{M}_{\alpha}} D_{\varphi}\left(\F_0^{-1},\left(\frac{1}{1-\lambda}\F_T - \frac{\lambda}{1-\lambda}\F_1(.|\theta)\right)^{-1}\right).\nonumber\\
\label{eqn:EstimProcLmomNotDual}
\end{eqnarray}
%where $\left(\frac{1}{1-\lambda}\F_T - \frac{\lambda}{1-\lambda}\F_1(.|\theta)\right)^{-1}$ purely denotes\footnote{We chose this notation to emphasize the link with mixture structure and its parameters.} the quantile measure associated to the cdf $\frac{1}{1-\lambda}\mathbb{F}_T - \frac{\lambda}{1-\lambda}\mathbb{F}_1(.|\theta)$. 
Now by virture of (\ref{eqn:F0InIntersect}), it holds that
\begin{equation}
(\lambda^*,\theta^*,\alpha^*) \in \arginf_{(\lambda,\theta,\alpha)\in\Phi^+}\inf_{\F_0^{-1}\in\mathcal{M}_{\alpha}} D_{\varphi}\left(\F_0^{-1},\left(\frac{1}{1-\lambda}\F_T - \frac{\lambda}{1-\lambda}\F_1(.|\theta)\right)^{-1}\right).
\label{eqn:EstimProcQuantileFuns}
\end{equation}
Next, we will treat this estimation procedure using the Fenchel duality in order to write a feasible optimization procedure, and then proceed to build upon a plug-in estimator based on an observed dataset $X_1,\cdots,X_n$.
%%%%%%%%%%%%%%%%%%%%%%%%%%%%%%%%%%%%%%%%%%%%%%%%%%%%%%%%%%%%%%%%%%%%%%%
\subsection{Estimation using the duality technique}
\begin{figure}[ht]
\centering
\includegraphics[scale=0.45]{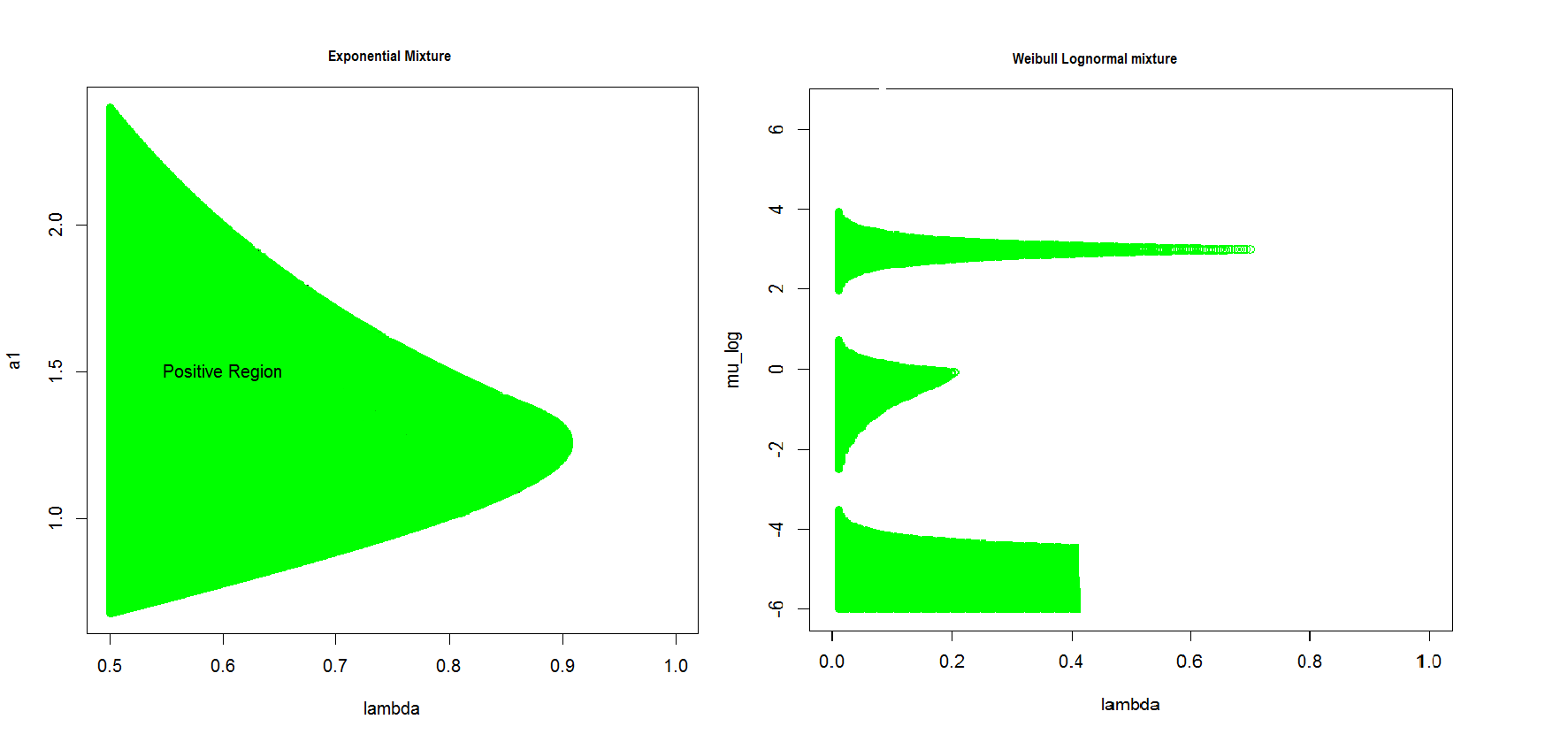}
\caption{Different forms of the set $\Phi^+$. For the Weibull-Lognormal mixture, the Weibull is the semiparametric component.}
\label{fig:DiffFormPhiPlus}
\end{figure}

Applying the duality result (\ref{eqn:DualityLmom}) on the estimation procedure (\ref{eqn:EstimProcLmomNotDual}) gives:
\[D_{\varphi}\left(\cup_{\alpha}\mathcal{M}_{\alpha}, \mathcal{N}^{-1}\right) =  \inf_{(\lambda,\theta,\alpha)\in\Phi^+}\sup_{\xi\in\mathbb{R}^{\ell-1}} \xi^t m(\alpha) - \int_0^1{\psi\left(\xi^tK(u)\right)d\left(\frac{1}{1-\lambda}\F_T - \frac{\lambda}{1-\lambda}\F_1(.|\theta)\right)^{-1}(u)}.\]
\noindent In order to keep formulas clearer, we adapt the following notation:
\[
\mathbb{F}_0(y|\phi) = \frac{1}{1-\lambda} \mathbb{F}_T(y) - \frac{\lambda}{1-\lambda} \mathbb{F}_1(y|\theta)
\]
Note that we must ensure the integrability condition 
\[\int{\|K\left(\mathbb{F}_0(y|\phi)\right)\|} dx <\infty,\]
in order to be able to use the duality technique. This is ensured by the definition of the polynomial vector $K$. Indeed, there exists a constant $c$ such that:
\[\left\|K\left(\mathbb{F}_0(y|\phi)\right)\right\| \leq c\left(\mathbb{F}_0(y|\phi)\right)\left(1-\mathbb{F}_0(y|\phi)\right).\]
Since $\mathbb{F}_0(y|\phi) = \left(\frac{1}{1-\lambda}\mathbb{F}_T(y) - \frac{\lambda}{1-\lambda}\mathbb{F}_1(y|\theta)\right)$ is supposed here to be a cdf because $(\lambda,\theta,\alpha)\in\Phi^+$, it suffices then that $\mathbb{F}_0(y|\phi)$ has a finite expectation so that the previous integral becomes finite.\\
This formulation is only useful when one has a sample of i.i.d. observations of the distribution $\frac{1}{1-\lambda}P_T - \frac{\lambda}{1-\lambda}P_1(.|\theta)$ for every $\lambda$ and $\theta$, because the integral can be approximated directly using the order statistics as in formula (\ref{eqn:DualityLmomEmpirical}). We need, however, a formula which shows directly the cdf because it would permit to approximate directly the objective function and avoid the calculus of the inverse of $\frac{1}{1-\lambda}\mathbb{F}_T - \frac{\lambda}{1-\lambda}\mathbb{F}_1(.|\theta)$. Besides, the replacement of the true cdf by the empirical one does not guarantee that the difference $\frac{1}{1-\lambda}\mathbb{F}_T - \frac{\lambda}{1-\lambda}\mathbb{F}_1(.|\theta)$ remains a cdf and more complications would appear in the proof of the consistency.\\ 
Using Lemma 1.2 from \cite{AlexisThesis} we may make the change of variable desired.
\[\int_0^1{\psi\left(\xi^tK(u)\right)d\left(\frac{1}{1-\lambda}\F_T - \frac{\lambda}{1-\lambda}\F_1(.|\theta)\right)^{-1}(u)} = \int_{\mathbb{R}}{\psi\left(\xi^tK\left(\mathbb{F}_0(y|\phi)\right)\right)dx}.\]
We have now:
\begin{equation}
(\lambda^*,\theta^*,\alpha^*) \in \arginf_{(\lambda,\theta,\alpha)\in\Phi^+}\sup_{\xi\in\mathbb{R}^{\ell-1}} \xi^tm(\alpha) - \int_{\mathbb{R}}{\psi\left(\xi^tK\left(\frac{1}{1-\lambda}\mathbb{F}_T(x) - \frac{\lambda}{1-\lambda}\mathbb{F}_1(x|\theta)\right)\right)dx}.
\label{eqn:EstimProcPhiPLus}
\end{equation}
We may now construct an estimator of $\phi^*$ by replacing $\mathbb{F}_T$ by the empirical cdf calculated on the basis of an i.i.d. sample $X_1,\cdots,X_n$. The resulting estimation procedure is still very complicated. This is mainly because we need to characterize the set $\Phi^+$. It is possible but is very expensive. For example, we may think about checking if the derivative with respect to $x$ $\frac{1}{1-\lambda}p_T(x) - \frac{\lambda}{1-\lambda}p_1(x|\theta)$ is nonnegative at a large randomly selected set of points. On the other hand, the set $\Phi^+$ can take \emph{fearful} forms for some mixtures. In Figure (\ref{fig:DiffFormPhiPlus}), we have two examples of $\Phi^+$. In the exponential mixture (the figure to the left), $\Phi^+$ has a "good" form in the sense that it is convex and contains $(\lambda^*,\theta^*)=(0.7,1.5)$ with a sufficiently large neighborhood around it. Thus, optimization procedures should not face any problem finding the optimum. However, in the Weibull-Lognormal mixture (the figure to the right) with $(\lambda^*,\mu^*)=(0.7,3)$, the set $\Phi^*$ is not even connected. Besides, there is not a sufficient neighborhood around $(\lambda^*,\mu^*)$ which permits an optimization algorithm to move around. During my simulations on data generated from a Weibull-Lognormal mixture distribution, the optimization algorithms could not reach such optimum and were always stuck at the initial point. A solution will be proposed in the next paragraph where we introduce the final step in the sequel of this estimation procedure.
%%%%%%%%%%%%%%%%%%%%%%%%%%%%%%%%%%%%%%%%%%%%%%%%%%%%%%%%%%%%%%%%%%%%%%
\subsection{The algorithm in practice and a plug-in estimate}
The problem with the estimation procedure (\ref{eqn:EstimProcPhiPLus}) is that the optimization is carried over the set $\Phi^+$ which may take "non-practical forms" as explained in the previous paragraph. The problem can be reread otherwise. The difficulty comes mainly from the fact that function $\frac{1}{1-\lambda}\mathbb{F}_T - \frac{\lambda}{1-\lambda}\mathbb{F}_1(.|\theta)$ may not be a cdf and the quantile would not exist. Thus, the estimation procedure in formula (\ref{eqn:EstimProcQuantileFuns}) cannot be used. We have, however, made disappear the quantiles in formula (\ref{eqn:EstimProcPhiPLus}) using a change of variable. Besides, there is no problem in calculating the optimized function in formula (\ref{eqn:EstimProcPhiPLus}) for any triplet $(\lambda,\theta,\alpha)\in\Phi$ even if the parameters do not define a proper cdf for function $\frac{1}{1-\lambda}\mathbb{F}_T - \frac{\lambda}{1-\lambda}\mathbb{F}_1(.|\theta)$. Besides, and more importantly, $\phi^*$ is a global infimum of the objective function $H(\phi,\xi(\phi))$ over the whole set $\Phi$ (and not only overy $\Phi^+$) where:
\[H(\phi,\xi)  =  \xi^t m(\alpha) - \int{\psi\left[\xi^tK\left(\frac{1}{1-\lambda} \mathbb{F}_T(y) - \frac{\lambda}{1-\lambda} \mathbb{F}_1(y|\theta)\right)\right]dy}\]
and $\xi(\phi) = \argsup_{\xi\in\mathbb{R}^{\ell-1}} H(\phi,\xi)$. Indeed, for any $\phi\in\Phi$, we have:
\[H(\phi,\xi(\phi)) \geq H(\phi,0) = 0.\]
Besides, using the duality attainment at $\phi=\phi^*$, we may write
\begin{eqnarray*}
H(\phi^*,\xi(\phi^*)) & = & \inf_{\F_0^{-1}\in\mathcal{M}_{\alpha^*}}D_{\varphi}\left(\F_0^{-1},\left(\frac{1}{1-\lambda^*}\F_T - \frac{\lambda^*}{1-\lambda^*}\F_1(.|\theta^*)\right)^{-1}\right) \\
 & = & \inf_{\F_0^{-1}\in\mathcal{M}_{\alpha^*}}D_{\varphi}\left(\F_0^{-1},{\F_0^*}^{-1}\right)\\
 & = & 0.
\end{eqnarray*}
Thus, if function $H(\phi,\xi(\phi))$ does not have several global infima inside $\Phi$, $(\lambda^*,\theta^*,\alpha^*)$ will hold as the only global minimum of it. In other words
\begin{equation}
\phi^* = \arginf_{(\alpha,\theta,\lambda)\in\Phi}\sup_{\xi\in\mathbb{R}^{\ell-1}} \xi^t m(\alpha) - \int{\psi\left[\xi^tK\left(\frac{1}{1-\lambda} \mathbb{F}_T(x) - \frac{\lambda}{1-\lambda} \mathbb{F}_1(x|\theta)\right)\right]dx}.
\label{eqn:EstimProcLmomDualCDFVersionTrue}
\end{equation}
Provided an i.i.d. sample $X_1,\cdots,X_n$ distributed from $P_T$, the cdf $\mathbb{F}_T$ can be approximated by its empirical version $\frac{1}{n}\sum{\ind{X_i\leq x}}$. Hence, $\phi^*$ can be estimated by:
\begin{equation}
\hat{\phi} = \arginf_{(\alpha,\theta,\lambda)\in\Phi}\sup_{\xi\in\mathbb{R}^l} \xi^t m(\alpha) - \int{\psi\left[\xi^tK\left(\frac{1}{1-\lambda} \mathbb{F}_n(x) - \frac{\lambda}{1-\lambda} \mathbb{F}_1(x|\theta)\right)\right]dx}.
\label{eqn:EstimProcLmomDualCDFVersion}
\end{equation}
\begin{remark}
Notice that the dual attainment no longer holds on the complementary set $\Phi\setminus\Phi^+$ since we are working with "signed cumulative functions". Our idea is to offer the optimization algorithm a larger neighborhood around the optimum in order to be able to find it. The important fact in the extended procedure is that $\phi^*$ is a \emph{global} infimum of the objective function. Our simulation study shows that the extension to $\Phi$ does not affect the results in several examples, and the estimator $\hat{\phi}$ is not biased and has an acceptable variance, see Section \ref{sec:SimulationLmom} for more details.
\end{remark}
%%%%%%%%%%%%%%%%%%%%%%%%%%%%%%%%%%%%%%%%%%%%%%%%%%%%%%%%%%%%%%%%%%%%%%%%%%%%%%%%
%%%%%%%%%%%%%%%%%%%%%%%%%%%%%%%%%%%%%%%%%%%%%%%%%%%%%%%%%%%%%%%%%%%%%%%%%%%%%%%%
\subsection{Uniqueness of the solution "under the model"}
By a unique solution we mean that only one quantile measure, which can be written in the form of $\left(\frac{1}{1-\lambda}F_T-\frac{\lambda}{1-\lambda}\F_1(.|\theta)\right)^{-1}$ for $(\lambda,\theta)\in\Phi^+$, verifies the L-moments constraints with a unique triplet $(\lambda^*,\theta^*,\alpha^*)$. The existence of a unique solution is essential in order to ensure that the procedure (\ref{eqn:EstimProcPhiPLus}) is a reasonable estimation method. We provide next a result ensuring the uniqueness of the solution. The proof is differed to Appendix \ref{AppendSemiPara:Prop1Lmom}. The proof does not provide sufficient conditions for the existence of a unique solution over $\Phi$ because in the proof we only study the intersection $\mathcal{N}^{-1}\cap\mathcal{M}$ and characterize it without using the Fenchel duality.
\begin{proposition}
\label{prop:UniqueSolLmom}
Assume that ${\F_0^*}^{-1}\in\mathcal{M}=\cup_{\alpha}\mathcal{M}_{\alpha}$. Suppose also that:
\begin{enumerate}
\item the system of equations:
\begin{equation}
\int_{0}^1{K(u)d\left(\frac{1}{1-\lambda}\F_T - \frac{\lambda}{1-\lambda}\F_1(.|\theta)\right)^{-1}(u)} = m(\alpha)
\label{eqn:NlnSysLmom}
\end{equation}
has a unique solution $(\lambda^*,\theta^*,\alpha^*)$;
\item the function $\alpha\mapsto m(\alpha)$ is one-to-one;
\item for any $\theta\in\Theta$ we have :
\[\lim_{x\rightarrow \infty} \frac{p_1(x|\theta)}{p_T(x)} = c,\quad \text{ with } c\in [0,\infty)\setminus\{1\};\]
\item the parametric component is identifiable, i.e. if $p_1(.|\theta) = p_1(.|\theta')\;\; dP_T-$a.e. then $\theta=\theta'$,
\end{enumerate}
then, the intersection $\mathcal{N}^{-1}\cap\mathcal{M}$ contains a unique measure ${\F_0^*}^{-1}$, and there exists a unique vector $(\lambda^*,\theta^*,\alpha^*)$ such that $P_T = \lambda^*P_1(.|\theta^*)+(1-\lambda)P_0^*$ where $P_0^*$ is defined through (\ref{eqn:TrueF0model}) and belongs to $\mathcal{M}_{\alpha^*}$. Moreover, provided assumptions 2-4, the conclusion holds if and only if assumption 1 is fulfilled.
\end{proposition}
There is no general result for a non linear system of equations to have a unique solution; still, it is necessary to ensure that we impose a number of constraints at least equal to the number of unknown variables, otherwise there would be an infinite number of $\sigma-$finite measures in the intersection $\mathcal{N}^{-1} \bigcap \cup_{\alpha\in\mathcal{A}}\mathcal{M}_{\alpha}$.
\begin{remark}
Assumptions 3 and 4 of Proposition \ref{prop:UniqueSolLmom} are used to prove the identifiability of the "model" $\left(\frac{1}{1-\lambda}P_T - \frac{\lambda}{1-\lambda}P_1(.|\theta)\right)_{\lambda,\theta}$. These conditions may be rewritten using the cdf. Furthermore, according to the considered situation we may find simpler ones for particular cases (or even for the general case). Our assumptions remain sufficient but not necessary for the proof. Note also that similar assumption to 3 can be found in the literature on semiparametric mixture models, see Proposition 3 in \cite{Bordes06b}. We can imagine the case of a semiparametric mixture model with a Gaussian parametric component with unknown mean. Besides, the unknown component could have a heavier tail such as an exponential model. Assumption 3 is then fulfilled with $c=0$.
\end{remark}

%%%%%%%%%%%%%%%%%%%%%%%%%%%%%%%%%%%%%%%%%%%%%%%%%%%%%%%%%%%%%%%%%%%%%%%%%%%%%%%%%%%%%%%%%%%%%%%%%%%%%%%
%
% ======================================================================
%%%%%%%%%%%%%%%%%%%%%%%%%%%%%%%%%%%%%%%%%%%%%%%%%%%%%%%%%%%%%%%%%%%%%%%%%%%%%%%%
% ======================================================================
%
%%%%%%%%%%%%%%%%%%%%%%%%%%%%%%%%%%%%%%%%%%%%%%%%%%%%%%%%%%%%%%%%%%%%%%%%%%%%%%%%%%%%%%%%%%%%%%%%%%%%%%%

\section{Asymptotic properties}\label{sec:AsymptoticProp}
We study the asymptotic properties of the estimator $\hat{\phi}$ defined by (\ref{eqn:EstimProcLmomDualCDFVersion}). For the consistency, we will assume that function $H(\phi,\xi(\phi))$ has a unique infimum on $\Phi$. This infimum is a fortiori $\phi^*$.
On the other hand, the limiting law would not change if the infimum is truly $\phi^*$ or any other point. $\hat{\phi}$ will be centered at the infimum with a multivariate Gaussian limit law. It would not, however, be interesting unless it is centered around $\phi^*$.
\subsection{Consistency}
We will use Theorem 4.1 from \cite{DiaaAssiaMoments} since we are in the same context of double optimization. We rewrite this theorem here in order to keep things clearer and for the sake of an exhaustive text. Suppose that our estimator $\hat{\phi}$ is defined through the following double optimization procedure. Let $H$ and $H_n$ be two generic functions such that $H_n(\phi,\xi) \rightarrow H(\phi,\xi)$ in probability for any couple $(\phi,\xi)$. Define $\hat{\phi}$ and $\phi^*$ as follows:
\begin{eqnarray*}
\hat{\phi} & = & \arginf_{\phi\in\Phi} \sup_{\xi\in\mathbb{R}^{\ell}} H_n(\phi,\xi);\\
\phi^* & = & \arginf_{\phi\in\Phi} \sup_{\xi\in\mathbb{R}^{\ell}} H(\phi,\xi).
\end{eqnarray*}
We adapt the following notation:
\[\xi(\phi) = \argsup_{t\in\mathbb{R}^{\ell}} H(\phi,t), \qquad \xi_n(\phi) = \argsup_{t\in\mathbb{R}^{\ell}} H_n(\phi,t)\]
The following theorem provides sufficient conditions for consistency of $\hat{\phi}$ towards $\phi^*$. This result will then be applied to the case of our estimator.\\
Assumptions:
\begin{itemize}
\item[A1.] the estimate $\hat{\phi}$ exists (even if it is not unique);
\item[A2.] $\sup_{\xi,\phi} \left|H_n(\phi,\xi) - H(\phi,\xi)\right|$ tends to 0 in probability;
\item[A3.] for any $\phi$, the supremum of $H$ over $\xi$ is unique and isolated, i.e. $\forall \varepsilon>0, \forall \tilde{\xi}$ such that $\|\tilde{\xi}-\xi(\phi)\|>\varepsilon$, then there exists $\eta>0$ such that $H(\phi,\xi(\phi)) - H(\phi,\tilde{\xi})>\eta$;
\item[A4.] the infimum of $\phi\mapsto H(\phi,\xi(\phi))$ is unique and isolated, i.e. $\forall \varepsilon>0, \forall \phi$ such that $\|\phi-\phi^*\|>\varepsilon$, there exists $\eta>0$ such that $H(\phi,\xi(\phi))-H(\phi^*,\xi(\phi^*))>\eta$;
\item[A5.] for any $\phi$ in $\Phi$, function $\xi\mapsto H(\phi,\xi)$ is continuous.
\end{itemize}
In assumption A4, we suppose the existence and uniqueness of $\phi^*$. It does not, however, imply the uniqueness of $\hat{\phi}$. This is not a problem for our consistency result. The vector $\hat{\phi}$ may be any point which verifies the minimum of function $\phi\mapsto\sup_{\xi} H_n(\phi,\xi)$. Our consistency result shows that all vectors verifying the minimum of $\phi\mapsto\sup_{\xi} H_n(\phi,\xi)$ converge to the unique vector $\phi^*$. We also prove an asymptotic normality result which shows that even if $\hat{\phi}$ is not unique, all possible values should be in a neighborhood of radius $\mathcal{O}(n^{-1/2})$ centered at $\phi^*$.\\
The following lemma establishes a uniform convergence result for the argument of the supremum over $\xi$ of function $H_n(\phi,\xi)$ towards the one of function $H(\phi,\xi)$. It constitutes a first step towards the proof of convergence of $\hat{\phi}$ towards $\phi^*$.
%%%%%%%%%%%%%%%%%%%%%%%%%%%
\begin{lemma}
\label{lem:SupXiPhiDiff}
Assume A2 and A3 are verified, then 
\[\sup_{\phi}\|\xi_n(\phi) - \xi(\phi)\| \rightarrow 0 ,\qquad \textit{in probability.}\] 
\end{lemma}
\noindent We proceed now to announce our consistency theorem. 
%%%%%%%%%%%%%%%%%%%%%%%%%
\begin{theorem}
\label{theo:MainTheorem}
Assume that A1-A5 are verified, then $\hat{\phi}$ tends to $\phi^*$ in probability.
\end{theorem}
The proofs of Lemma \ref{lem:SupXiPhiDiff} and Theorem \ref{theo:MainTheorem} can be found in \cite{DiaaAssiaMoments}. Let's start by precising the functions $H$ and $H_n$.
\begin{eqnarray*} 
H(\phi,\xi) & = & \xi^t m(\alpha) - \int{\psi\left[\xi^tK\left(\frac{1}{1-\lambda} \mathbb{F}_T(y) - \frac{\lambda}{1-\lambda} \mathbb{F}_1(y|\theta)\right)\right]dy}; \\
H_n(\phi,\xi) & = & \xi^t m(\alpha) - \int{\psi\left[\xi^tK\left(\frac{1}{1-\lambda} \mathbb{F}_n(y) - \frac{\lambda}{1-\lambda} \mathbb{F}_1(y|\theta)\right)\right]dy},
\end{eqnarray*}
and recall the notations:
\begin{eqnarray*}
\mathbb{F}_0(y|\phi) & = & \frac{1}{1-\lambda} \mathbb{F}_T(y) - \frac{\lambda}{1-\lambda} \mathbb{F}_1(y|\theta); \\
\hat{\mathbb{F}}_0(y|\phi) & = & \frac{1}{1-\lambda} \mathbb{F}_n(y) - \frac{\lambda}{1-\lambda} \mathbb{F}_1(y|\theta);\\
\xi(\phi) & = & \argsup_{\xi\in\mathbb{R}^{\ell-1}} H(\phi,\xi); \\
\xi_n(\phi) & = & \argsup_{\xi\in\mathbb{R}^{\ell-1}} H_n(\phi,\xi).
\end{eqnarray*}
We start by calculating the difference $H(\phi,\psi) - H_n(\phi,\psi)$.
\begin{equation}
H(\phi,\xi) - H_n(\phi,\xi) = \int{\psi\left[\xi^tK\left(\hat{\mathbb{F}}_0(y|\phi)\right)\right] - \psi\left[\xi^tK\left(\mathbb{F}_0(y|\phi)\right)\right] dy}.
\label{eqn:DiffFunConsistency}
\end{equation}
The following lemma is essential for the proof of the consistency. We need to transform the optimization over $\xi$ onto a compact set. Thus, \emph{important} values of $\xi$ which are necessary for the calculus of the supremum are bounded. The proof is differed to Appendix \ref{Append:LemmaCompactXi}.
%%%%%%%%%%%%%%%%%%%%%%%%%%%%%%%%%%%%%%%%%%%%
\begin{lemma}
\label{lem:LmomCompactXi}
Suppose that function $\xi\mapsto H(\phi,\xi)$ is of class $\mathcal{C}^2(\mathbb{R}^{\ell-1})$. Then, functions $\phi\mapsto\xi(\phi)$ and $\phi\mapsto\xi_n(\phi)$ are well defined and $\mathcal{C}^1$ on the interior of the whole set $\Phi$. Moreover, if $\Phi$ is compact, then $\hat{\phi}$ and $\phi^*$ exist and the sets Im$(\xi(.))$ and Im$(\xi_n(.))$ are compact.
\end{lemma}

\noindent Differentiability of function $H$ with respect to $\xi$ can be checked in general using Lebesgue theorems, but it would not have been wise to impose an assumption over the integrand since $\psi'$ is increasing and $\xi$ is a priori in $\mathbb{R}^{\ell-1}$. For the class of functions of Cressie-Read (\ref{eqn:CressieReadPhi}), we have $\psi(t)=\frac{1}{\gamma}(\gamma t - t +1)^{\gamma/(\gamma-1)}-\frac{1}{\gamma}$. Thus, for $\gamma>1$, $\psi(t)\rightarrow\infty$ as $t\rightarrow\infty$. Therefore, it is important to study each special case alone. For example, $\psi(y)=y^2/2+y$ is the dual of the Chi square generator $\varphi(t)=(t-1)^2/2$, then $H(\xi,\phi)$ is a polynomial of degree 2 in $\xi$ and hence differentiable up to second order, see Example \ref{Example:Chi2Lmom} below for more details.\\
We state the consistency of the estimator $\hat{\phi}$ defined by (\ref{eqn:EstimProcLmomDualCDFVersion}). The proof is based on Theorem \ref{theo:MainTheorem} and is differed to Appendix \ref{Append:TheoConsistLmom}.
%%%%%%%%%%%%%%%%%%%%%%%%
\begin{theorem}
\label{theo:ConsistencyLmom}
Suppose that
\begin{itemize}
\item[C1.] $\Phi$ is a compact subset of $\mathbb{R}^d$;
\item[C2.] function $\psi$ is continuously differentiable;
\item[C3.] the infimum of $\phi\mapsto H(\phi,\xi(\phi))$ is unique and isolated, i.e. $\forall \varepsilon>0, \forall \phi$ such that $\|\phi-\phi^*\|>\varepsilon$, there exists $\eta>0$ such that $H(\phi,\xi(\phi))-H(\phi^*,\xi(\phi^*))>\eta$;
\item[C4.] function $\alpha\mapsto m(\alpha)$ is continuous;
\item[C5.] there exists an integrable function $\hat{h}$ such that $\left|\psi\left[\xi^tK\left(\mathbb{F}_0(y,|\phi)\right)\right]\right|<\hat{h}(y)$;
\item[C6.] the integral $\int{\sqrt{\mathbb{F}_T(y)(1-\mathbb{F}_T(y))}dy}$ is finite,
\end{itemize}
then the estimator $\hat{\phi}$ defined by (\ref{eqn:EstimProcLmomDualCDFVersion}) converges in probability to $\phi^*$.
\end{theorem}
%%%%%%%%%%%%%%%%%%%%%%%
\begin{remark}
If we use $\tilde{\phi}$ defined by (\ref{eqn:EstimProcPhiPLus}), only assumption C3 should be changed. We need to suppose that the infimum exists and is unique inside $\Phi^+$ instead of the whole parameter space $\Phi$. This is less restrictive than assumption C3 since we are working inside a subset of $\Phi$.
\end{remark}
\begin{remark}
Assumption C5 is only used to prove continuity of function $H(\phi,\xi)$ with respect to $\xi$ which we discussed after Lemma \ref{lem:LmomCompactXi}. When $\psi(t)=t^2/2+t$, then assumption C5 is not fulfilled, but function $H(\phi,\xi)$ is still continuous as a function of $\xi$, because it is a polynomial of order 2 in $\xi$. Assumption C6 will be needed again in the proof of the asymptotic normality. Sufficient conditions are discussed in Remark \ref{remark:RegVar} hereafter.
\end{remark}
\begin{example}[$\chi^2$ case]
\label{Example:Chi2Lmom}
The case of the $\chi^2$ divergence is very interesting, because the optimization over $\xi$ can be calculated. Write function $H(\phi,\xi)$ for $\psi(t)=t^2/2+t$.
\[H(\phi,\xi) = \xi^tm(\alpha) - \int{\frac{1}{2}\left(\xi^tK\left(\mathbb{F}_0(y|\phi)\right)\right)^2 + \xi^tK\left(\mathbb{F}_0(y|\phi)\right) dy}.\]
This is a polynomial of order 2 in $\xi$ and thus $H(\phi,\xi)$ is of class $\mathcal{C}^2(\mathbb{R}^{\ell-1})$ as soon as the integrals exist. Indeed, for any $r\leq\ell$, there exists $c_r$ such that:
\begin{eqnarray}
\left|K_r\left(\mathbb{F}_0(y,|\phi)\right) \right| & \leq & c_r \left|\mathbb{F}_0(y,|\phi)\left(1-\mathbb{F}_0(y,|\phi)\right)\right|\label{eqn:IntegrabilityKF0Part1} \\
 & \leq & \frac{c_r}{(1-\lambda)^2}\left[\mathbb{F}_T(y)\left(1-\mathbb{F}_T(y)\right) + \lambda\mathbb{F}_T(y)(1-\mathbb{F}_1(y))+\lambda\mathbb{F}_1(1-\mathbb{F}_T(y))+ \right.\nonumber \\
 & & \left. \lambda^2\mathbb{F}_1(y)(1-\mathbb{F}_1(y))\right]. \label{eqn:IntegrabilityKF0Part2}
\end{eqnarray}
For example, if the distributions $\mathbb{F}_T$ and $\mathbb{F}_1$ are defined on $\mathbb{R}_+$, then the right hand side is integrable as soon as the expectations of $\mathbb{F}_T$ and $\mathbb{F}_1$ are finite.\\
A simple calculus of the derivative of function $\xi\mapsto H(\phi,\xi)$ gives
\[
\frac{\partial H}{\partial \xi} (\xi,\phi) = m(\alpha) - \int{K\left(\mathbb{F}_0(y,|\phi)\right)\xi^tK\left(\mathbb{F}_0(y,|\phi)\right)dy} + \int{K\left(\mathbb{F}_0(y,|\phi)\right)dy}.
\]
The optimum is attained for:
\[\xi(\phi) = \Omega^{-1}\left(m(\alpha) - \int{K(\mathbb{F}_0(y|\phi))}dy\right),\]
where
\[\Omega = \int{K\left(\mathbb{F}_0(y|\phi)\right)K\left(\mathbb{F}_0(y|\phi)\right)^tdy}.\]
Furthermore, the Hessian matrix is equal to $-\Omega$, so it is symmetric definite negative whatever the value of the vector $\phi$. Thus, $\xi(\phi)$ is a global maximum of function $\xi\mapsto H(\xi,\phi)$ for any $\phi\in\Phi$. This was not the case for moment-type constraints since the Hessian matrix might be definite positive for some values of the vector $\phi$, see \cite{DiaaAssiaMoments}Example 4.1. The empirical version of this calculus is obtained similarly by replacing $\mathbb{F}_0(y|\phi)$ by $\hat{\mathbb{F}}_0(y|\phi)$.\\
Conditions of the consistency theorem can be verified. Assumption C1 is very natural in practice since in general, we have in mind a range of values for the parameters. Assumption C2 is fulfilled since $\psi(t)$ is polynomial of degree 2. Assumption 3 is not simple in general and depends on the model. Assumption C4 follows the problem we have. In Example \ref{Example:Weibull}, $m(\alpha) = (-\lambda_2,-\lambda_3,-\lambda_4)$ is continuous on $(0,\infty)\times(0,\infty)$, and assumption C4 becomes verified. We have verified assumption C5 at the beginning of the example. Assumption C6 is not restrictive. It is verified for example in an exponential mixture. The idea is to control the tail behavior of the distribution, see remark (\ref{remark:RegVar}) for a general approach.
\end{example}

%==========================================================
%%%%%%%%%%%%%%%%%%%%%%%%%%%%%%%%%%%%%%%%%%%%
\subsection{Asymptotic normality}
The convergence in law of the estimator $\hat{\phi}$ defined by (\ref{eqn:EstimProcLmomDualCDFVersion}) is not simply deduced in the same way we obtained it in the moment-constraints case (\cite{DiaaAssiaMoments} Section 4.2). A Taylor expansion would not show directly the empirical distribution which combined with the CLT gives the asymptotic normality. The expansion results in the term $\int{K(\hat{\mathbb{F}}_0(x))dx}$ which is a functional of the empirical distribution. In the case of SPLQ models (no parametric component) presented in paragraph (\ref{subsec:SPLQDef}), \cite{AlexisGSI13} used a result based on Theorem 6 from \cite{Stigler} to study the quantity which corresponds to $\int{K(\hat{\mathbb{F}}_0(x))dx}$. This result is based on sums of order statistics which \emph{cannot} be adapted to our context since $\hat{\mathbb{F}}_0$ is an estimator of $\mathbb{F}_0$ different from the corresponding empirical distribution, besides $\hat{\mathbb{F}}_0$ may not be a "proper" cdf. We present here a result for our situation. Our proof still shares a part of the idea of the proof of the result of \cite{Stigler}, see Appendix \ref{Append:PropAsyptotNormIntegral}. This constitutes a first step in the proof of the asymptotic normality.\\
\begin{proposition}
\label{prop:LimitLawLmomConstrPart}
Suppose that $\mathbb{E}|X_i|<\infty$. Suppose also that
\begin{eqnarray}
\int{\sqrt{\mathbb{F}_T(y)\left(1-\mathbb{F}_T(y)\right)}dy} & < & \infty, \label{eqn:AsymptotNormCond1}\\
\int{\int{\mathbb{F}_T(\min(x,y)) - \mathbb{F}_T(x)\mathbb{F}_T(y)dx}dy} & < & \infty. \label{eqn:AsymptotNormCond2}
\end{eqnarray}
For any vector $\phi=(\lambda,\theta,\alpha)\in\Phi$, we then have
\begin{multline*}
\sqrt{n}\left[\int{K\left(\frac{1}{1-\lambda}\mathbb{F}_n(y) - \frac{\lambda}{1-\lambda}\mathbb{F}_1(y|\theta)\right)dy} - \int{K\left(\frac{1}{1-\lambda}\mathbb{F}_T(y) - \frac{\lambda}{1-\lambda}\mathbb{F}_1(y|\theta)\right)dy}\right] \\ \xrightarrow{\quad \mathcal{D} \quad}{}\mathcal{N}\left(0,\Sigma\right), 
\end{multline*}
where the covariance matrix $\Sigma$ is given by
\begin{equation}
\Sigma_{r_1,r_2} = \int{\int{\left(\mathbb{F}_T\left(\min(x,y)\right) - \mathbb{F}_T(x)\mathbb{F}_T(y)\right)\sum_{k=0}^{r_1-1}{c_{r_1,k}\mathbb{F}_0(x|\phi)^k}\sum_{k=0}^{r_2-1}{c_{r_2,k}\mathbb{F}_0(y|\phi)^k}dy}dx},
\label{eqn:VarCovMatConstrPart}
\end{equation}
and $c_{r,k}=(-1)^{r-k-1}\binom{r-1}{k}\binom{r+k-1}{k}$ for $r,r_1,r_2\in\{2,\cdots,\ell\}$.
\end{proposition}
\begin{remark}
It was not possible to use a functional delta method (see \cite{Vaart} Chap. 20, Theorem 20.8) to prove the limiting law here because the functional $G\mapsto\int{K(\xi^tK(G))}$ is not Hadamard differentiable.
\end{remark}
\begin{remark}
\label{remark:RegVar}
Integrability conditions (\ref{eqn:AsymptotNormCond1}) and (\ref{eqn:AsymptotNormCond2}) over the distribution function can be reformulated by imposing directly conditions over the distribution function using the notion of regular variations and the Lemma page 280 in \cite{Feller}. Regular variations transform the problem into conditions over the tails of the distribution functions. Suppose that there exists a constant $\rho_+<-2$ and a function $L_+(x)$ such that:
\begin{equation}
1-\mathbb{F}_T(x) = x^{\rho_+}L_+(x), \text{  with }\qquad \frac{L_+(tx)}{L_+(t)}\xrightarrow[\quad t\rightarrow\infty\quad]{} 1,\forall x>0.
\label{eqn:AsymptotNormCond1Alter1}
\end{equation}
Then, the integral $\int_y^{\infty}{\sqrt{1-\mathbb{F}_T(x)}dx}$ converges and there exists a function $M_+(y)$ such that $M_+(ty)/M_+(t)\rightarrow 1,\forall y$ and $\int_y^{\infty}{\left[1-\mathbb{F}_T(x)\right]dx} = y^{\rho_+ +1}M_+(y)$. For the neighborhood of $-\infty$, we make similar assumptions over $\mathbb{F}_T(x)$. Suppose that there exists a constant $\rho_-<-2$ and a function $L_-(x)$ such that:
\begin{equation}
\mathbb{F}_T(x) = x^{\rho_-}L_-(x), \text{  with } \qquad \frac{L_-(-tx)}{L_-(t)}\xrightarrow[\quad t\rightarrow-\infty\quad]{} 1,\forall x<0.
\label{eqn:AsymptotNormCond1Alter2}
\end{equation}
Then, the integral $\int_{-\infty}^y{\sqrt{\mathbb{F}_T(x)}dx}$ converges and there exists a function $M_-(y)$ such that $M_-(ty)/M_-(t)\rightarrow 1,\forall y$ and $\int_y^{\infty}{\mathbb{F}_T(x)dx} = y^{\rho_- +1}M_-(y)$.\\
These two assertions permit to conclude that condition (\ref{eqn:AsymptotNormCond1}) is verified since $\sqrt{\mathbb{F}_T(x)(1-\mathbb{F}_T(x))}\leq \sqrt{\mathbb{F}_T(x)}\ind{x\in(-\infty,0)} + \sqrt{1-\mathbb{F}_T(x)}\ind{x\in(0,\infty)}$. Moreover, condition (\ref{eqn:AsymptotNormCond2}) can also be check. Let's discuss what happens when $y$ is at a neighborhood of either $+\infty$ or $-\infty$. For any $y>0$, one may write:
\begin{eqnarray*}
\int_y^{+\infty}{[\mathbb{F}_T(\min(x,y))-\mathbb{F}_T(y)\mathbb{F}_T(x)]dx} & = & \mathbb{F}_T(y)\int_y^{+\infty}{[1-\mathbb{F}_T(x)]dx} \\
 & = & \mathbb{F}_T(y) y^{\rho_+ +1}M_+(y)
\end{eqnarray*}
which is integrable in a neighborhood of $+\infty$ with respect to $y$ by (\ref{eqn:AsymptotNormCond1Alter1}). On the other hand, for any $y<0$, one may write
\begin{eqnarray*}
\int_{-\infty}^y{[\mathbb{F}_T(\min(x,y))-\mathbb{F}_T(y)\mathbb{F}_T(x)]dx} & = & \left[1-\mathbb{F}_T(y)\right]\int_{-\infty}^y{\mathbb{F}_T(x)dx} \\
 & = & \left[1-\mathbb{F}_T(y)\right] y^{\rho_- +1}M_-(y)
\end{eqnarray*}
which is integrable in a neighborhood of $-\infty$ with respect to $y$ by (\ref{eqn:AsymptotNormCond1Alter2}). Thus, condition (\ref{eqn:AsymptotNormCond2}) is ensured under assumptions (\ref{eqn:AsymptotNormCond1Alter1},\ref{eqn:AsymptotNormCond1Alter2}).\\
\end{remark}
%%%%%%%%%%%%%%%%%%%%%%%%%%%%
\noindent We move on now to show the asymptotic normality of the estimator $\hat{\phi}$. Define the following matrices:
\begin{eqnarray}
J_{\phi^*,\xi^*} & = & \left(\begin{array}{c}-\int{\left[\frac{1}{(1-\lambda^*)^2}\mathbb{F}_T(y) - \frac{1}{(1-\lambda^*)^2}\mathbb{F}_1(y|\theta^*)\right]K'(\mathbb{F}_0(y|\phi^*))dy} \\ \frac{\lambda^*}{1-\lambda^*} \int{\nabla_{\theta}\mathbb{F}_1(y|\theta^*)K'(\mathbb{F}_0(y|\phi^*))^tdy} \\ \nabla m(\alpha^*) \end{array}\right)^t\label{eqn:NormalAsymLMomJ1} \\
J_{\xi^*,\xi^*} & = & \int{K(\mathbb{F}_0(y|\phi^*))K(\mathbb{F}_0(y|\phi^*))^tdy} \label{eqn:NormalAsymLMomJ2} \\
\tilde{\Sigma} & = & \left(J_{\phi^*,\xi^*}^t J_{\xi^*,\xi^*} J_{\phi^*,\xi^*}\right)^{-1} \\  
H & = & \tilde{\Sigma} J_{\phi^*,\xi^*}^t J_{\xi^*,\xi^*}^{-1} \label{eqn:NormalAsymLmomH}\\
P & = & J_{\xi^*,\xi^*}^{-1} - J_{\xi^*,\xi^*}^{-1} J_{\phi^*,\xi^*} \tilde{\Sigma} J_{\phi^*,\xi^*}^t J_{\xi^*,\xi^*}^{-1} \label{eqn:NormalAsymLmomP}
\end{eqnarray}
We use the same notations considered at the beginning of this section for $\mathbb{F}_0(x|\phi),\hat{\mathbb{F}}_0(x|\phi),\xi(\phi)$ and $\xi_n(\phi)$.
\begin{theorem}
\label{theo:AsymptotNormLmom}
Suppose that assumptions of Proposition \ref{prop:LimitLawLmomConstrPart} are fulfilled. Suppose also that
\begin{enumerate}
\item $(\hat{\phi},\xi_n(\hat{\phi}))$ tends to $(\phi^*,0)$ in probability;
\item $\phi^*\in$ int$(\Phi)$;
\item $\alpha\mapsto m(\alpha)$ is of class $\mathcal{C}^2$;
\item there exist an integrable function $B_1$ such that $\|\nabla_{\theta}\mathbb{F}_1(y|\theta)\|\leq B_1(y)$ for $\theta$ in a neighborhood of $\theta^*$;
\item there exist integrable functions $B_{2,1}$ and $B_{2,2}$ such that $\|\nabla_{\theta}\mathbb{F}_1(y|\theta)\nabla_{\theta}\mathbb{F}_1(y|\theta)^t\|\leq B_{2,2}(y)$ and $\|J_{\mathbb{F}_1(y|\theta)}\|\leq B_{2,1}(y)$ for $\theta$ in a neighborhood of $\theta^*$;
\item the integral $\int{[\mathbb{F}_T(y)-\mathbb{F}_1(y)]dy}$ exists and is finite;
\item the matrices $J_{\xi^*,\xi^*}$ and $J_{\phi^*,\xi^*}^t J_{\xi^*,\xi^*} J_{\phi^*,\xi^*}$ are invertible.
\end{enumerate}
Then,
\[\left(\begin{array}{c}  \sqrt{n}\left(\hat{\phi}-\phi^*\right) \\ \sqrt{n}\xi_n(\hat{\phi})\end{array}\right) \xrightarrow[\mathcal{L}]{} \mathcal{N}\left(0,\left(\begin{array}{c}H \\ P\end{array}\right) \Sigma \left(H^t\quad P^t\right)\right),\]
where $H,P$ and $\Sigma$ are given respectively by formulas (\ref{eqn:NormalAsymLmomH}), (\ref{eqn:NormalAsymLmomP}) and (\ref{eqn:VarCovMatConstrPart}).
\end{theorem}
\noindent The proof of this theorem is differed to Appendix \ref{Append:TheoNormalAsymptotLmom}. In assumption 1, we could only demand the consistency of $\hat{\phi}$, since the consistency of $\xi_n(\hat{\phi})$ can be deduced from it using the continuity of $\phi\mapsto\xi(\phi)$, see Lemma \ref{lem:LmomCompactXi}, and the uniform convergence of $\xi_n(.)$ towards $\xi(.)$, see Lemma \ref{lem:SupXiPhiDiff}. Assumptions 4-6 are used in the proof to ensure the differentiability up to second order with respect to $\xi$ and $\phi$ of $H_n(\phi,\xi)$ for any $n$.
%%%%%%%%%%%%%%%%%%%%%%%%%%%%%%%%%%%%%%%%%%%%%%%%%%%%%%%%%%%%%%%%%%%%%
% ===============================================
%%%%%%%%%%%%%%%%%%%%%%%%%%%%%%%%%%%%%%%%%%%%%%%%%%%%%%%%%%%%%%%%%%%%
\section{Simulation study}\label{sec:SimulationLmom}
We perform several simulations and show how a prior information about the distribution of the semiparametric component $P_0$ can help us better estimate the set of parameters $(\lambda^*,\theta^*,\alpha^*)$ in regular examples, i.e. the components of the mixture can be clearly distinguished when we plot the probability density function. We also show how our approach permits to estimate even in difficult situations when the proportion of the parametric component is very low; such cases could \emph{not} be handled using existing methods. We show also the advantage of using L-moments constraints over moment constraints using the approach developed in \cite{DiaaAssiaMoments} especially when the proportion of the parametric component is very low. \\
In our experiments, the datasets were generated by the following mixtures:
\begin{itemize}
\item[$\bullet$] A two-component Weibull mixture;
\item[$\bullet$] A two-component Weibull -- Lognormal mixture;
\item[$\bullet$] A two-component Gaussian -- Two-sided Weibull mixture;
\end{itemize}
We have chosen a variety of values for the parameters especially the proportion. We only used the $\chi^2$ divergence, because the optimization over $\xi$ can be calculated without numerical methods, see Example \ref{Example:Chi2Lmom} above and Example 4.1 from \cite{DiaaAssiaMoments}. Since the objective function $\phi\mapsto H_n(\phi,\xi_n(\phi))$ as a function of $\phi$ is not ensured to be strictly convex, we used 6 fixed initial points which we specify for each example separately. We then ran the Nelder-Mead algorithm and chose the vector of parameters for which the objective function has the lowest value. We applied a similar procedure on the algorithm of \cite{Bordes10} in order to ensure a \emph{fair} comparison. For the moment-type method of \cite{DiaaAssiaMoments}, we used 10 random initial points in the effective parameter space (inside the set where the optimized function over $\xi$ is concave).\\
All numerical integrations were calculated using function \texttt{integral} of package \texttt{pracma}. It performed better than the standard function \texttt{integrate} and the function \texttt{distrExIntegrate} of package \texttt{distrEx}.\\
We did not use any error criterion function (such as the total variation distance) here because the compared methods do not provide the same set of parameters. For example, the method of \cite{Bordes10} estimates a mean value for the unknown component whereas our approach estimates a shape parameter. Other existing methods do not estimate any information about the parameters of the unknown component.\\
We did not use the EM-type methods of \cite{Robin} and \cite{BordesStochEM} when the number of observations is greater than $10^4$, because of the high execution time they need. Besides, their performance in lower number of observation suffices to do the comparison with new procedure.

\begin{remark}
For the methods of \cite{Song}, we need to estimate mixture's distribution using a kernel density estimator. For the data generated from a Weibull mixture and the data generated from a Weibull--Lognormal mixture, we used a reciprocal inverse Gaussian kernel density estimator with a window equal to 0.01 according to the simulation study in \cite{Diaa}. For the method of \cite{Bordes10}, we used a triangular kernel which gave better results than the use of a Gaussian kernel.
\end{remark}
\begin{remark}
In the literature on the stochastic EM algorithm, it is advised that we iterate the algorithm for some time until it reaches a stable state, then continue iterating long enough and average the values obtained in the second part. The trajectories of the algorithm were very erratic especially for the estimation of the proportion. For us, we iterated for the stochastic EM-type algorithm of \cite{BordesStochEM} 5000 times and averaged the 4000 final iterations.
\end{remark}
\begin{remark}
Initialization of both the EM-type algorithm of \cite{Song} and the SEM-type algorithm of \cite{BordesStochEM} was not very important, and we got the same results when the vector of weights was initialized uniformly or in a "good" way. The method of \cite{Robin} was more influenced by such initialization and we used most of the time a good starting points. In the paper of \cite{Robin}, the authors mention that their EM-type algorithm has a fixed point with a proportion at 0 or 1. This confirms that a good initialization is needed in order to avoid theses extreme solutions.
\end{remark}

%%%%%%%%%%%%%%%%%%%%%%%%%%%%%%%%%%%%
\subsection{Data generated from a two-component Weibull mixture modeled by a semiparametric Weibull mixture}
We consider a mixture of two Weibull components with scales $\sigma_1 = 0.5,\sigma_2=1$ and shapes $\nu_1=2,\nu_2=1$ in order to generate the dataset. In the semiparametric mixture model, the parametric component will be "the one to the right", i.e. the component whose true set of parameters is $(\nu_1=2,\sigma_1=0.5)$.\\
We impose on the unknown component three L-moments constraints; the second, the third and the fourth Weibull L-moments. They are given in Example \ref{Example:Weibull}. This mixture was not easily estimtated by either our estimation procedure or the semiparametric methods from the literature. Our estimator, although has a higher variance, is still not biased in the same way estimates of other methods are. The L-moment constraints gave an estimator with less variance than the estimator based on moments constraints, but with slightly higher bias on the proportion.
\begin{table}[ht]
\centering
\begin{tabular}{|c|c|c|c|c|c|c|}
\hline
Nb of observations & $\lambda$ & sd$(\lambda)$ & $\nu_1$ & sd($\nu_1$) & $\nu_2$ & sd($\nu_2$)\\
\hline
\hline
\multicolumn{7}{|c|}{Mixture 1 : $n=10^4$ $\lambda^* = 0.3$, $\nu_1^*=2$, $\sigma_1^*=0.5$(fixed), $\nu_2^*=1$, $\sigma_2^*=1$(fixed) }\\
\hline
Pearson's $\chi^2$ 3 moments & 0.304 & 0.016 & 2.191 & 0.887 & 0.998 & 0.013 \\
Pearson's $\chi^2$ 3 L-moments & 0.348 & 0.062 & 1.828 & 0.648 & 0.984 & 0.021 \\
Robin & 0.604 & 0.029 & 1.256  & 0.037 & --- & --- \\
Song EM-type & 0.806 & 0.005 & 1.185 & 0.018 & --- & --- \\
Song $\pi-$maximizing & 0.624 & 0.007 & 1.312 & 0.013 & --- & --- \\
\hline
\end{tabular}
\caption{The mean value with the standard deviation of estimates in a 100-run experiment on a two-component Weibull mixture.}
\label{tab:3by3ResultsWeibullLMoment}
\end{table}

%%%%%%%%%%%%%%%%%%%%%%%%%%%%%%%%%%%%%%%%%

\subsection{Data generated from a two-component Weibull-LogNormal mixture modeled by a semiparametric Weibull-LogNormal mixture}
We consider a dataset generated from a mixture of a Weibull and a Lognormal distributions. The Weibull component has a scale $\sigma_1^*=1$ and a shape $\nu_1^*\in\{1.5,1,0.4\}$ in order to illustrate several scenarios; a distribution whose pdf explodes to infinity at zero, a distribution whose pdf has finite value at zero and a distribution whose pdf goes back to zero at zero. The Lognormal component has a scale $\sigma_2^*=0.5$ and a mean parameter $\mu^*=3$. The Lognormal distribution has a heavy tail which is inherited in the mixture distribution.\\
In a first part, we perform a comparison of convergence speed between the method under moments constraints and the method under L-moments constraints as we increase the number of observations $n$. The Weibull component is considered as the unknown component during estimation, and is defined by its three L-moments constraints; the second, the third and the fourth. The first 4 L-moments of the Weibull distribution are given in Example \ref{Example:Weibull}.\\
In a second part, we perform an estimation of a semiparametric mixture model where the Lognormal component is considered unknown and defined through 3 L-moments conditions; the second, the third and the fourth L-moment.  The L-moments of the Lognormal distribution do not have a close formula and are calculated numerically using function \texttt{lmrln3} of package \texttt{lmom} written by Hosking.\\
Results in table (\ref{tab:3by3ResultsWeibullLognormLmom}) show that L-moments are more informative and we need less data in order to get good estimates in comparison to moments constraints. In order to calculate the estimate $\hat{\phi}$, we considered 6 initial points; namely the set 
\[\phi^{(0)} \in \left\{(0.8,2,1),(0.5,2,1),(0.8,1,1),(0.7,3,1.5),(0.7,2,2),(0.5,4,2),(0.5,1.5,2)\right\}.\] 
The vector $\hat{\phi}$ was taken as the one which corresponds to the lowest value among the infima produced by the optimization algorithm.\\
In table (\ref{tab:3by3ResultsLognormWeibullLmoment}) the Lognormal component is the unknown component during estimation. Initialization of the optimization algorithm, for example in mixture 2, was taken from the set $\{(0.1,0.5,1),(0.15,0.5,0.7),(0.05,1.5,2.5),(0.1,1,3)\}$.\\
It is clear that the moments constraints gave better results than L-moments constraints in mixture 1 for the estimation of the scale of the Weibull component. For the second mixture, both types of constraints give similar results. The two methods have the same bias in the estimation of the scale of Weibull component; the moments constraints produced a positive bias whereas the L-moments constraints produced a negative bias. The L-moments produced a smaller variance. In the third mixture, the L-moments constraints gave clear better results. The last mixture is the most difficult one in the sense that the proportion of the parametric component is very low. 

\begin{table}[ht]
\centering
\begin{tabular}{|c|c|c|c|c|c|c|c|}
\hline
nb of observations & Estimation method & $\lambda$ & sd$(\lambda)$ & $\mu$ & sd($\mu$) & $\nu$ & sd($\nu$)\\
\hline
\hline
\multicolumn{8}{|c|}{True Parameters : $\lambda^* = 0.7$, $\mu^*=3$, $\sigma_2^*=0.5$(fixed), $\nu^*=1.5$, $\sigma_1^*=1$(fixed) }\\
\hline
\multirow{2}{2.5cm}{$n = 10^2$} & L-moments & 0.685 & 0.069 & 2.798 & 0.413 & 0.436 & 0.074 \\
 & Moments & 0.384 & 0.117 & 2.654 & 0.153 & 0.488 & 0.018 \\
\hline 
\multirow{2}{2.5cm}{$n=10^3$} & L-moments & 0.677 & 0.017 & 3.014 & 0.028 & 0.726 & 0.272 \\
 & Moments & 0.518 & 0.068 & 2.806 & 0.099 & 0.473 & 0.014 \\
\hline
\multirow{2}{2.5cm}{$n=10^4$} & L-moments & 0.697 & 0.009 & 3.003 & 0.010 & 1.343 & 0.185 \\
 & Moments & 0.605 & 0.044 & 2.903 & 0.069 & 0.531 & 0.326 \\
\hline
\end{tabular}
\caption{The mean value with the standard deviation of estimates in a 100-run experiment on a two-component Weibull-log normal mixture.}
\label{tab:3by3ResultsWeibullLognormLmom}
\end{table}

\begin{table}[ht]
\centering
\begin{tabular}{|c|c|c|c|c|c|c|}
\hline
Nb of observations & $\lambda$ & sd$(\lambda)$ & $\nu$ & sd($\nu$) & $\mu$ & sd($\mu$)\\
\hline
\hline
\multicolumn{7}{|c|}{Mixture 1 : $n=10^3$, $\lambda^* = 0.3$, $\nu^*=1.5$, $\sigma_1^*=1$(fixed), $\mu^*=3$, $\sigma_2^*=0.5$(fixed) }\\
\hline
Pearson's $\chi^2$ L-moments& 0.313 & 0.019 & 1.027 & 0.541 & 2.992 & 0.050 \\
Pearson's $\chi^2$ Moments& 0.308 & 0.017 & 1.484 & 0.624 & 3.002 & 0.026 \\
Robin & 0.296 & 0.015 & 1.557 & 0.068 & --- & --- \\
Song EM-type & 0.291 & 0.015 & 1.614 & 0.087 & --- & --- \\
Song $\pi-$maximizing & 0.230 & 0.022 & 1.662 & 0.251 & --- & --- \\
SEM & 0.284 & 0.041 & 1.570 & 0.263 & --- & ---\\
\hline
\hline
\multicolumn{7}{|c|}{Mixture 2 : $n=10^4$, $\lambda^* = 0.1$, $\nu^*=1$, $\sigma_1^*=1$(fixed), $\mu^*=3$, $\sigma_2^*=0.5$(fixed) }\\
\hline
Pearson's $\chi^2$ L-moments & 0.104 & 0.006 & 0.795 & 0.379 & 2.994 & 0.015 \\
Pearson's $\chi^2$ Moments & 0.103 & 0.006 & 1.284 & 0.677 & 3.001 & 0.007 \\
Robin & 0.095 & 0.003 & 1.049 & 0.031 & --- & --- \\
Song EM-type & 0.100 & 0.004 & 0.894 & 0.039 & --- & --- \\
Song $\pi-$maximizing & 0.085 & 0.005 & 1.024 & 0.055 & --- & --- \\
SEM & 0.094 & 0.015 & 1.054 & 0.228 & --- & --- \\
%\hline
%\hline
%\multicolumn{7}{|c|}{Mixture 3 : $n=10^4$, $\lambda^* = 0.05$, $\nu^*=1$, $\sigma_1^*=1$(fixed), $\mu^*=3$, $\sigma_2^*=0.5$(fixed) }\\
%\hline
%Pearson's $\chi^2$ & 0.052 & 0.004 & 1.312 & 0.703 & 3.001 & 0.006 \\
\hline
\hline
\multicolumn{7}{|c|}{Mixture 3 : $n=5\times 10^4$, $\lambda^* = 0.05$, $\nu^*=0.4$, $\sigma_1^*=1$(fixed), $\mu^*=3$, $\sigma_2^*=0.5$(fixed) }\\
\hline
Pearson's $\chi^2$ L-Moments & 0.049 & 0.002 & 0.448 & 0.129 & 3.000 & 0.006 \\
Pearson's $\chi^2$ Moments & 0.049 & 0.002 & 0.629 & 0.438 & 3.001 & 0.004 \\
Song EM-type & 0.064 & 0.001 & 0.345 & 0.004 & --- & --- \\
Song $\pi-$maximizing & 0.024 & 0.001 & 0.773 & 0.010 & --- & --- \\
\hline
\end{tabular}
\caption{The mean value with the standard deviation of estimates in a 100-run experiment on a two-component Weibull-log normal mixture.}
\label{tab:3by3ResultsLognormWeibullLmoment}
\end{table}

%%%%%%%%%%%%%%%%%%%%%%%%%%%%%%%%%%%%%%%%%

\subsection{Data generated from a two-sided Weibull Gaussian mixture modeled by a semiparametric two-sided Weibull Gaussian mixture}
The (symmetric) two-sided Weibull distribution can be considered as a generalization of the Laplace distribution and can be defined through either its density or its distribution function as follows:
\[f(x|\nu,\sigma) = \frac{1}{2}\frac{\sigma}{\nu}\left(\frac{|x|}{\sigma}\right)^{\nu-1}e^{-\left(\frac{|x|}{\sigma}\right)^{\nu}}, \qquad \mathbb{F}(x|\nu,\sigma) = \left\{ \begin{array}{cc} 1-\frac{1}{2}e^{-\left(\frac{x}{\sigma}\right)^{\nu}} & x\geq 0 \\ 
e^{-\left(\frac{-x}{\sigma}\right)^{\nu}} & x< 0\end{array} \right.\]
We can also define a skewed form of the two-sided Weibull distribution by attributing different scale and shape parameters to the positive and the negative parts, and then normalizing in a suitable way so that $f(x)$ integrates to one; see \cite{Chen2sideWeibull}.
The 2nd, 3rd and 4th L-moments of the two-sided Weibull distribution are given by:
\begin{eqnarray*}
\lambda_2 & = & \left[1-\frac{1}{2^{1+1/\nu}}\right]\sigma_2\Gamma\left(1+\frac{1}{\nu}\right);\\
\lambda_3 & = & 0 ;\\
\lambda_4 & = & \left[1-\frac{6}{2^{1+1/\nu}}+\frac{15}{2\times 3^{1+1/\nu}}-\frac{5}{2\times 4^{1+1/\nu}}\right] \sigma_2\Gamma\left(1+\frac{1}{\nu}\right).
\end{eqnarray*}
\noindent We simulate different samples from a two-component mixture with a parametric component $f_1$ a Gaussian $\mathcal{N}(\mu=0,\sigma=0.5)$ and a semiparametric component $f_0$ a (symmetric) two-sided Weibull distribution with parameters $\nu\in\{3,1.5\}$ and a scale $\sigma_0\in\{1.5,2\}$. We perform different experiments to estimate the proportion and the mean of the parametric part (the Gaussian) and the shape of the semiparametric component. The values of the scale of the two components are considered to be known during estimation.\\
Results are presented in table (\ref{tab:3by3ResultsTwoSideWeibullGaussLMom}). The L-moments constraints produce clear better results than the moments constraints in all the mixtures. The estimation based on L-moments constraints produced clear lower variance. Besides, and once again, the L-moments constraints seem to be more informative and we need less number of observations than moments constraints in order to produce good estimates.\\
In this example we presented a challenge to our estimation method by simulating mixtures with very low proportion of the parametric part; mixture 3 with $\lambda^*=0.05$ and mixture 4 with $\lambda^*=0.01$. Using signal-noise terms, in mixture 4, only one percent of the data comes from the signal whereas $99\%$ of the data is pure noise. The location of the signal is then estimated around zero with standard deviation of $0.3$ with the L-moments constraints. It is not well localized however using moments constraints with $10^5$ observations, and we need at least $10^8$ observations to reach a similar precision to the result obtained with L-moments constraints. It is still important to notice that using moments or L-moments constraints, we were able to confirm the existence of a signal component (the parametric component).\\
In what concerns the initialization of the algorithm under L-moments constraints, we used:
\begin{eqnarray*}
\text{Mix 1} & : & \left\{(0.8,1,1),(0.5,-1,2.5),(0.8,0.5,2),(0.7,0,3),(0.7,1,4),(0.5,2,3.5)\right\} \\
\text{Mix 2} & : & \left\{(0.2,1,1),(0.5,-1,2.5),(0.2,0.5,2),(0.3,0,3),(0.3,1,4)\right\} \\
\text{Mix 3} & : & \left\{(0.1,1,1),(0.05,-1,2.5),(0.03,0.5,2),(0.01,0,1.5),(0.005,1,0.7)\right\} \\
 \text{Mix 4} & : & \left\{(0.1,1,1),(0.005,1,0.7)\right\}
\end{eqnarray*}
For the last mixture, we have found no changes in using more initial points than the two given points. Besides, execution time was very long (about 5 samples per day), so we preferred to use only two starting points.
\begin{table}[ht]
\centering
\begin{tabular}{|c|c|c|c|c|c|c|}
\hline
Estimation method & $\lambda$ & sd$(\lambda)$ & $\mu$ & sd($\mu$) & $\nu$ & sd($\nu$)\\
\hline
\hline
\multicolumn{7}{|c|}{Mixture 1 : $n=100$, $\lambda^* = 0.7$, $\mu^*=0$, $\sigma_2^*=0.5$(fixed), $\nu^*=3$, $\sigma_1^*=1.5$(fixed) }\\
\hline
Pearson's $\chi^2$ -- L-Moments & 0.758 & 0.067 & -2.28$\times 10^{-3}$ & 0.098 & 3.040  & 0.639 \\
Pearson's $\chi^2$ under $\mathcal{M}_{2:4}$ & 0.764 & 0.067 & -0.012 & 0.342 & 2.893  & 0.731 \\
Bordes symmetry Triangular Kernel & 0.309 & 0.226 & 0.240 & 0.609 & $\mu_2=-$0.220 & sd$(\mu_2)$0.398 \\
Robin et al. & 0.488 & 0.137 & -0.005 & 0.114 & --- & --- \\
EM-type Song et al. & 0.762 & 0.040 & -0.005 & 0.092 & --- & --- \\
$\pi-$maximizing Song et al. & 0.717 & 0.156 & -0.161 & 2.301 & --- & --- \\
Stochastic EM & 0.539 & 0.083 & -0.005 & 0.112 & --- & --- \\
\hline
\hline
\multicolumn{7}{|c|}{Mixture 2 : $n=100$, $\lambda^* = 0.3$, $\mu^*=0$, $\sigma_2^*=0.5$(fixed), $\nu^*=3$, $\sigma_1^*=1.5$(fixed) }\\
\hline
Pearson's $\chi^2$ -- L-Moments & 0.364 & 0.082 & -0.016 & 0.246 & 3.058  & 0.418 \\
Pearson's $\chi^2$ under $\mathcal{M}_{2:4}$ & 0.407 & 0.077 & 0.012 & 0.575 & 2.925  & 0.454 \\
Bordes symmetry Triangular Kernel & 0.272 & 0.119 & 0.773 & 0.947 & $\mu_2=-$0.430 & sd$(\mu_2)=$0.393 \\
%Bordes symmetry Gaussian Kernel & 0.206 & 0.104 & 0.855 & 0.911 & $\mu_2=-$0.308 & sd$(\mu_2)=$0.350 \\
Robin et al. & 0.203 & 0.078 & -0.109 & 0.947 & --- & --- \\
EM-type Song et al. & 0.494 & 0.035 & -0.132 & 0.806 & --- & --- \\
$\pi-$maximizing Song et al. & 0.384 & 0.129 & 0.014 & 1.321 & --- & --- \\
Stochastic EM & 0.263 & 0.040 & -0.062 & 0.646 & --- & --- \\
\hline
\hline
\multicolumn{7}{|c|}{Mixture 3 : $n=5000$, $\lambda^* = 0.05$, $\mu^*=0$, $\sigma_2^*=0.5$(fixed), $\nu^*=1.5$, $\sigma_1^*=2$(fixed) }\\
\hline
Pearson's $\chi^2$ -- L-Moments & 0.050 & 0.013 & 0.026 & 0.365 & 1.496  & 0.020 \\
Pearson's $\chi^2$ under $\mathcal{M}_{2:4}$ 0.066 & 0.013 & -0.036 & 0.857 & 1.493 & 0.008\\
Robin et al. & 0.078 & 0.012 & -0.009 & 1.046 & --- & --- \\
EM-type Song et al. & 0.306 & 0.006 & -0.050 & 0.995 & --- & --- \\
$\pi-$maximizing Song et al. & 0.001 & 0.001 & -0.546 & 2.314 & --- & --- \\
\hline
\hline
\multicolumn{7}{|c|}{Mixture 4 : $n=10^5$, $\lambda^* = 0.01$, $\mu^*=0$, $\sigma_2^*=0.5$(fixed), $\nu^*=1.5$, $\sigma_1^*=2$(fixed) }\\
\hline
Pearson's $\chi^2$ -- L-Moments & 0.011 & 0.003 & 0.023 & 0.377 & 1.500  & 0.005 \\
Pearson's $\chi^2$ under $\mathcal{M}_{2:4}$ & 0.025 & 0.010&  - 0.047 & 1.356 & 1.495 & 0.006\\
\hline
\end{tabular}
\caption{The mean value with the standard deviation of estimates in a 100-run experiment on a two-component two-sided Weibull--Gaussian mixture under L-moment constraints.}
\label{tab:3by3ResultsTwoSideWeibullGaussLMom}
\end{table}

\subsection{Conclusions}
In this paper, we introduced a new structure for semiparametric mixture models with unknown component by imposing L-moments constraints on it. The resulting estimators were proved to be consistent and asymptotically normal under standard assumptions. The estimation method under L-moments constraints presented several advantages in comparison to the estimation method under moments constraints in \cite{DiaaAssiaMoments}. We were able to estimate over the whole parameter space $\Phi$ instead of only a subset $\Phi^+$ of it, and no need to check if the optimized function $\xi\mapsto H(\phi,\xi)$ is strictly concave for every $\phi$. Although the estimation method under L-moments constraints need numerical integrations (which is not the case of moments-type constraints procedure), the resulting estimators seem to have lower variance in general. Moreover, L-moments are demonstrated through simulations to be more informative than moments constraints on several dataset, and we need less number of observations in order to obtain good estimates. A comparison on real data problem is required in order to confirm the importance of these approaches and will be considered in a future work.
\clearpage
%%%%%%%%%%%%%%%%%%%%%%%%%%%%%%%%%%%%%%%%%%%%%%%%%%%%%%%%%%%%%%%%%%%%%%%%%%%%%%%%%%%%%%%%%%%%%%%%%%%%%%%
%
% ======================================================================
%%%%%%%%%%%%%%%%%%%%%%%%%%%%%%%%%%%%%%%%%%%%%%%%%%%%%%%%%%%%%%%%%%%%%%%%%%%%%%%%
% ======================================================================
%
%%%%%%%%%%%%%%%%%%%%%%%%%%%%%%%%%%%%%%%%%%%%%%%%%%%%%%%%%%%%%%%%%%%%%%%%%%%%%%%%%%%%%%%%%%%%%%%%%%%%%%%

\section{Appendix: Proofs}
\subsection{Proof of Proposition \ref{prop:identifiabilityMixtureLmom}}\label{AppendSemiPara:PropIdenitifiabilityLmom}
\begin{proof}
Denote $M^{1}$ the set of all probability measures. Based on equation (\ref{eqn:IdenitifiabilityDefEqLmom}), we have:
\begin{eqnarray*}
P_0 & = & \frac{1}{1-\lambda} P_T - \frac{\lambda}{1-\lambda}P_1(.|\theta) \\
\tilde{P}_0 & = & \frac{1}{1-\tilde{\lambda}} P_T - \frac{\tilde{\lambda}}{1-\tilde{\lambda}}P_1(.|\tilde{\theta})
\end{eqnarray*}
Define the following function:
\[G:\mathbb{R}^{d-s}\times M^+\rightarrow \text{Im}(G)\subset M^1: (\lambda,\theta,P_0)\mapsto \lambda P_1(.|\theta) + (1-\lambda)P_0.\]
where 
\[M^+ = \{P_0 \in M^1 \text{  s.t.} \F_0^{-1}\in\mathcal{M}\}.\]
Identifiability is now equivalent to the fact that function $G$ is one-to-one. This means that for a given mixture distribution $P_T\in$Im$(G)$, we need that there exists a unique triplet $(\lambda,\theta,P_0)$ such that
\[P_T = \lambda P_1(.|\theta) + (1-\lambda)P_0\]
In other words:
\[P_0 = \frac{1}{1-\lambda}P_T - \frac{\lambda}{1-\lambda}P_1(.|\theta)\]
The equality of measures imply the equality of the quantiles. Thus, we may write:
\begin{equation}
\int_0^1{K(u)d\F_0^{-1}(u)} = m(\alpha) = \int_0^1{K(u)d\left(\frac{1}{1-\lambda}\F_T - \frac{\lambda}{1-\lambda}\F_1(.|\theta)\right)^{-1}(u)}
\label{eqn:SysLmom}
\end{equation}
The assumption of the present proposition imposes the existence of unique solution $(\lambda^*,\theta^*,\alpha^*)$ to the previous nonlinear system of equations. Let's go back to function $G$. For a given mixture distribution $P_T\in$Im$(G)$, take $\lambda=\lambda^*,\theta=\theta^*$ to be the solution to the nonlinear system (\ref{eqn:SysLmom}),  and define $P_0^*$ by:
\[P_0^* = \frac{1}{1-\lambda^*}P_T - \frac{\lambda^*}{1-\lambda^*}P_1(.|\theta^*).\]
Notice that $P_0^*\in\mathcal{M}_{\alpha^*}$. Suppose that $P_T$ can be written in two manners. In other words, suppose that there exists another triplet $(\tilde{\lambda},\tilde{\theta},\tilde{P}_0)$ with $\tilde{P}_0\in\mathcal{M}_{\tilde{\alpha}}$ such that:
\[P_T = \tilde{\lambda} P_1(.|\tilde{\theta}) + (1-\tilde{\lambda})\tilde{P}_0.\]
We then have:
\[\tilde{P}_0 = \frac{1}{1-\tilde{\lambda}}P_T - \frac{\tilde{\lambda}}{1-\tilde{\lambda}}P_1(.|\tilde{\theta}),\]
and consequently,
\[m(\tilde{\alpha}) = \int_0^1{K(u)d\left(\frac{1}{1-\tilde{\lambda}}\F_T - \frac{\tilde{\lambda}}{1-\tilde{\lambda}}\F_1(.|\tilde{\theta})\right)^{-1}(u)}.\]
Thus, $(\tilde{\lambda},\tilde{\theta},\tilde{\alpha})$ is a second solution to the system (\ref{eqn:SysLmom}). Nevertheless, the system of equations (\ref{eqn:SysLmom}) has a unique solution by assumption of the present proposition. Hence, a contradiction is reached and the triplet $(\lambda^*,\theta^*,P_0^*)$ is unique. We conclude that function $G$ is one-to-one and the semiparametric mixture model subject to L-moments constraints is identifiable.
\end{proof}

%%%%%%%%%%%%%%%%%%%%%%%%%%%%%%%%%%%%%%%%%%%%%%%%%%%%%%%
%%%%%%%%%%%%%%%%%%%%%%%%%%%%%%%%%%%%%%%%%%%%%%%%%%%%%%%

\subsection{Proof of Proposition \ref{prop:UniqueSolLmom}}\label{AppendSemiPara:Prop1Lmom}
\begin{proof}
Let $\F_0^{-1}$ be some quantile measure which belongs to the intersection $\mathcal{N}^{-1} \cap \mathcal{M}$. Since $\F_0^{-1}$ belongs to $\mathcal{N}^{-1}$, there exists a couple $(\lambda,\theta)\in\Phi^+$ such that:
\begin{equation}
\F_0^{-1} = \left(\frac{1}{1-\lambda} \F_T - \frac{\lambda}{1-\lambda} \F_1(.|\theta)\right)^{-1}.
\label{eqn:SetNelementQuantile}
\end{equation}
This couple is unique by virtue of assumptions 3 and 4. Indeed, let $(\lambda,\theta)$	and $(\tilde{\lambda},\tilde{\theta})$ be two couples such that:
\begin{equation*}
\left(\frac{1}{1-\lambda} \F_T - \frac{\lambda}{1-\lambda} \F_1(.|\theta)\right)^{-1} = \left(\frac{1}{1-\tilde{\lambda}} \F_T - \frac{\tilde{\lambda}}{1-\tilde{\lambda}} \F_1(.|\tilde{\theta})\right)^{-1}
\end{equation*}
This entails that:
\begin{equation}
\frac{1}{1-\lambda} \mathbb{F}_T(x) - \frac{\lambda}{1-\lambda} \mathbb{F}_1(x|\theta) = \frac{1}{1-\tilde{\lambda}} \mathbb{F}_T(x) - \frac{\tilde{\lambda}}{1-\tilde{\lambda}} \mathbb{F}_1(x|\tilde{\theta}).
\label{eqn:identifEqualityQuantile}
\end{equation}
By derivation of both sides, we get an identity in the densities:
\[\frac{1}{1-\lambda} - \frac{\lambda}{1-\lambda} \frac{p_1(x|\theta)}{p_T(x)} = \frac{1}{1-\tilde{\lambda}} - \frac{\tilde{\lambda}}{1-\tilde{\lambda}} \frac{p_1(x|\tilde{\theta})}{p_T(x)}.\]
Taking the limit as $x$ tends to $\infty$ results in:
\[\frac{1-c\lambda}{1-\lambda}  = \frac{1-c\tilde{\lambda}}{1-\tilde{\lambda}}.\]
Note that function $z\mapsto (1-cz)/(1-z)$ is strictly monotone as long as $c\neq 1$. Hence, it is a one-to-one map. Thus $\lambda=\tilde{\lambda}$. Inserting this result in equation (\ref{eqn:identifEqualityQuantile}) entails that:
\[\mathbb{F}_1(.|\theta) = \mathbb{F}_1(.|\tilde{\theta}).\]
Using the identifiability of $P_1$ (assumption 4), we get $\theta=\tilde{\theta}$ which proves the existence of a unique couple $(\lambda,\theta)$ in (\ref{eqn:SetNelementQuantile}).\\
On the other hand, since $\F_0^{-1}$ belongs to $\mathcal{M}$, there exists a unique $\alpha$ such that $\F_0^{-1}\in\mathcal{M}_{\alpha}$. Uniqueness comes from the fact that the function $\alpha\mapsto m(\alpha)$ is one-to-one (assumption 2). Thus, $\F_0^{-1}$ verifies the constraints
\[\int_0^1{K(u)d\F_0^{-1}(u)} = m(\alpha).\]
Combining this with (\ref{eqn:SetNelementQuantile}), we get:
\begin{equation}
\int_0^1{K(u)d\left(\frac{1}{1-\lambda} \F_T - \frac{\lambda}{1-\lambda} \F_1(.|\theta)\right)^{-1}(u)} = m(\alpha).
\label{eqn:NlnSysMalphaQuantile}
\end{equation}
This is a non linear system of equations with $\ell$ equations. Now, let $\F_0^{-1}$ and $\tilde{\F}_0^{-1}$ be two elements in $\mathcal{N}^{-1}\cap\mathcal{M}$, then there exist two couples $(\lambda,\theta)$ and $(\tilde{\lambda},\tilde{\theta})$ with $\lambda\neq\tilde{\lambda}$ or $\theta\neq\tilde{\theta}$ such that $\F_0^{-1}$ and $\tilde{\F}_0^{-1}$ can be written in the form of (\ref{eqn:SetNelementQuantile}) with respectively $(\lambda,\theta)$ and $(\tilde{\lambda},\tilde{\theta})$. Since $\F_0^{-1}\in\mathcal{M}$, there exists $\alpha$ such that $\F_0^{-1}\in\mathcal{M}_{\alpha}$. Similarly, there exists $\tilde{\alpha}$ possibly different from $\alpha$ such that $\tilde{\F}_0^{-1}\in\mathcal{M}_{\tilde{\alpha}}$. Now, $(\lambda,\theta,\alpha)$ and $(\tilde{\lambda},\tilde{\theta},\tilde{\alpha})$ are two solutions to the system of equations (\ref{eqn:NlnSysMalphaQuantile}) which contradicts with assumption 1 of the present proposition.\\
We may now conclude that, if a quantile measure $\F_0^{-1}$ belongs to the intersection $\mathcal{N}^{-1} \cap \mathcal{M}$, then it has the representation (\ref{eqn:SetNelementQuantile}) for a unique couple $(\lambda,\theta)$ and there exists a unique $\alpha$ such that the triplet $(\lambda,\theta,\alpha)$ is a solution to the non linear system (\ref{eqn:NlnSysMalphaQuantile}). Conversely, if there exists a triplet $(\lambda,\theta,\alpha)$ which solves the non linear system (\ref{eqn:NlnSysMalphaQuantile}), then the quantile measure $\F_0^{-1}$ defined by $\F_0^{-1} = \left(\frac{1}{1-\lambda} \F_T - \frac{\lambda}{1-\lambda} \F_1(.|\theta)\right)^{-1}$ belongs to the intersection $\mathcal{N}^{-1} \cap \mathcal{M}$. This is because on the one hand, it clearly belongs to $\mathcal{N}^{-1}$ by its definition and on the other hand, it belongs to $\mathcal{M}_{\alpha}$ since it verifies the constraints and thus belongs to $\mathcal{M}$.\\
It is now reasonable to conclude that under assumptions 2-4, the intersection $\mathcal{N}^{-1} \cap \mathcal{M}$ includes a \emph{unique} quantile measure $\F_0^{-1}$ if and only if the set of $\ell$ non linear equations (\ref{eqn:NlnSysMalphaQuantile}) has a unique solution $(\lambda,\theta,\alpha)$.
\end{proof}

%%%%%%%%%%%%%%%%%%%%%%%%%%%%%%%%%%%%%%%%%%%%%%%%%%%%%%%%%%%%%%%%%%%%%%%%%%%%%%%%%%%%%%%%%%%%%%%%%%%%%%%
%%%%%%%%%%%%%%%%%%%%%%%%%%%%%%%%%%%%%%%%%%%%%%%%%%%%%%%%%%%%%%%%%%%%%%%%%%%%%%%%%%%%%%%%%%%%%%%%%%%%%%%

\subsection{Proof of Lemma \ref{lem:LmomCompactXi}}\label{Append:LemmaCompactXi}
\begin{proof}
The same arguments hold for both functions $\xi(\phi)$ and $\xi_n(\phi)$. We therefore, proceed with $\xi(\phi)$. Function $\xi\mapsto H(\phi,\xi)$ is strictly concave since\footnote{One can prove the strict concavity simply by calculating $H(\phi,u\xi_1+(1-u)\xi_2)$.} it is $\mathcal{C}^2$ and have the following Hessian matrix:
\[J_{H(\phi,.)} = -\int{K\left(\mathbb{F}_0(y,|\phi)\right)K\left(\mathbb{F}_0(y,|\phi)\right)^t} \psi''\left(\xi^tK\left(\mathbb{F}_0(y,|\phi)\right)\right)dy.\]
Since $\psi$ is strictly convex, then $\psi''(z)>0$ for any $z$. Thus the matrix $J_{H(\phi,.)}$ is definite negative and $\xi\mapsto H(\phi,\xi)$ is strictly concave. By the implicit function theorem, function $\phi\mapsto\xi(\phi)$ is uniquely defined and $\mathcal{C}^1$ over int$(\Phi)$. Notice here that even if $\frac{1}{1-\lambda} \mathbb{F}_T(y) - \frac{\lambda}{1-\lambda} \mathbb{F}_1(y|\theta)$ is negative, the matrix $J_{H(\phi,.)}$ can still be definite negative unlike the case of moment constraints.\\
The second part of the proposition is a direct consequence from the continuity of function $\phi\mapsto\xi(\phi)$.
\end{proof}

%%%%%%%%%%%%%%%%%%%%%%%%%%%%%%%%%%%%%%%%%%%%%%%%%%%%%%%%%%%%%%%%%%%%%%%%%%%%%%%%%%%%%%%%%%%%%%%%%%%%%%%
%%%%%%%%%%%%%%%%%%%%%%%%%%%%%%%%%%%%%%%%%%%%%%%%%%%%%%%%%%%%%%%%%%%%%%%%%%%%%%%%%%%%%%%%%%%%%%%%%%%%%%%
\subsection{Proof of Theorem \ref{theo:ConsistencyLmom}}\label{Append:TheoConsistLmom}
\begin{proof}
We will use Theorem \ref{theo:MainTheorem}. We start with assumption A2. We prove, first, that the supremum over $\xi$ can only be calculated over a compact subset of $\mathbb{R}^l$. This is a direct result from Lemma \ref{lem:LmomCompactXi}. One can redefine the estimator by maximizing over $\xi$ on the subset $\Xi=$Im$(\xi(.))\subset\mathbb{R}^l$ independently of $\phi$. We thus have:
\begin{eqnarray*}
D_{\varphi}(\mathcal{M}_{\alpha},\mathbb{F}_0(.|\phi)) & = & \sup_{\xi\in\Xi} H(\phi,\xi) \\
\phi^* & = & \arginf_{\phi}\sup_{\xi\in\Xi} H(\phi,\xi).
\end{eqnarray*}
We redefine now the estimation procedure (\ref{eqn:EstimProcLmomDualCDFVersion}) as follows:
\[\hat{\phi} = \arginf_{\alpha,\theta,\lambda}\sup_{\xi\in\Xi} \xi^t m(\alpha) - \int{\psi\left[\xi^tK\left(\frac{1}{1-\lambda} \mathbb{F}_T(y) - \frac{\lambda}{1-\lambda} \mathbb{F}_1(y|\theta)\right)\right]dy}\]
Using the mean value theorem, there exists $\eta(y)\in(0,1)$ such that\footnote{In the case of the Chi square, $\lambda(y)=\frac{1}{2}$}:
\begin{multline}
\psi\left(\xi^tK\left(\mathbb{F}_0(y|\phi)\right)\right) - \psi\left(\xi^tK\left(\hat{\mathbb{F}}_0(y|\phi)\right)\right) = \xi^t\left(K\left(\mathbb{F}_0(y|\phi)\right) - K\left(\hat{\mathbb{F}}_0(y|\phi)\right)\right) \\ 
\times \psi'\left(\eta(y)\xi^tK\left(\mathbb{F}_0(y|\phi)\right) + (1-\eta(y))\xi^tK\left(\hat{\mathbb{F}}_0(y|\phi)\right)\right)
\label{eqn:MeanValResConsist}
\end{multline}
An exact formula of function $\eta(y)$ will not be needed. We will only use the fact that its image is included in $(0,1)$. By the central limit theorem, one can write:
\[\sqrt{n}\frac{\mathbb{F}_n(y) - \mathbb{F}_T(y)}{\sqrt{\mathbb{F}_T(y)(1-\mathbb{F}_T(y))}} \rightarrow \mathcal{N}\left(0,1\right).\]
Since $\hat{\mathbb{F}}_0(y|\phi) - \mathbb{F}_0(y|\phi) = \mathbb{F}_n(y) - \mathbb{F}_T(y)$, we write
\[\sqrt{n}\frac{\hat{\mathbb{F}}_0(y|\phi) - \mathbb{F}_0(y|\phi)}{\sqrt{\mathbb{F}_T(y)(1-\mathbb{F}_T(y))}} \rightarrow \mathcal{N}\left(0,1\right),\]
which entails by the delta method that:
\begin{equation}
\sqrt{n}\frac{K\left(\hat{\mathbb{F}}_0(y|\phi)\right) - K\left(\mathbb{F}_0(y|\phi)\right)}{\sqrt{\mathbb{F}_T(y)(1-\mathbb{F}_T(y))}} \rightarrow \mathcal{N}\left(0,\nabla K\left(\mathbb{F}_0(y|\phi)\right) \nabla K\left(\mathbb{F}_0(y|\phi)\right)^t\right).
\label{eqn:LimiLawConsist}
\end{equation}
Since function $K$ is a vector of polynomials, its gradient is a matrix of polynomials. Besides, the distribution function $\mathbb{F}_0(y|\phi)$ takes its values in $[0,1]$, thus the variance of the limiting law in (\ref{eqn:LimiLawConsist}) is of order $\frac{1}{n}$ independently of $y$ and $\phi$. We may now write:
\begin{equation}
\frac{K\left(\hat{\mathbb{F}}_0(y|\phi)\right) - K\left(\mathbb{F}_0(y|\phi)\right)}{\sqrt{\mathbb{F}_T(y)(1-\mathbb{F}_T(y))}} = o_P(1)
\label{eqn:KdiffConsist}
\end{equation}
Going back to equation (\ref{eqn:DiffFunConsistency}), we use equations (\ref{eqn:MeanValResConsist}) and (\ref{eqn:KdiffConsist}) to write:
\begin{eqnarray*}
H(\phi,\xi) - H_n(\phi,\xi) & = & \int{\xi^t\left(K\left(\mathbb{F}_0(y|\phi)\right) - K\left(\hat{\mathbb{F}}_0(y|\phi)\right)\right)\psi'\left[\eta(y)\xi^tK\left(\mathbb{F}_0(y|\phi)\right) \right.}\\
				& & \text{\vspace{2cm}} \left.+ (1-\eta(y))\xi^tK(\hat{\mathbb{F}}_0(y|\phi))\right]dy \\
 & = & \int{\sqrt{\mathbb{F}_T(y)(1-\mathbb{F}_T(y))}\xi^t\frac{\left(K\left(\mathbb{F}_0(y|\phi)\right) - K\left(\hat{\mathbb{F}}_0(y|\phi)\right)\right)}{\sqrt{\mathbb{F}_T(y)(1-\mathbb{F}_T(y))}} }\\
& & \times \psi'\left(\eta(y)\xi^tK\left(\mathbb{F}_0(y|\phi)\right) + (1-\eta(y))\xi^tK\left(\hat{\mathbb{F}}_0(y|\phi)\right)\right)dy \\
& = & \xi^to_p(1)\int{\sqrt{\mathbb{F}_T(y)(1-\mathbb{F}_T(y))}\psi'\left[\eta(y)\xi^tK\left(\mathbb{F}_0(y|\phi)\right)\right.}\\
& & \left. + (1-\eta(y))\xi^tK(\hat{\mathbb{F}}_0(y|\phi))\right]dy.
\end{eqnarray*}
The finale line can also be justified by the Chebyshev's inequality, see Remark \ref{rem:ConsistKdiffChebyshev}, or even using the calculus in the proof of Proposition \ref{prop:LimitLawLmomConstrPart} below.\\
It suffices now to prove that the integral in the previous display is finite. Here, $\xi$ (resp. $\phi$) is inside the compact set $\Xi$ (resp. $\Phi$), and functions $\eta(y), \mathbb{F}_0(y|\phi)$ and $\hat{\mathbb{F}}_0(y|\phi)$ all take values inside the compact interval $[0,1]$. Thus, continuity of $\psi'$ suffices to conclude that there exists a constant $M$ independent of $y$, $\phi$ and $\xi$ such that:
\begin{equation}
\left|\psi'\left(\eta(y)\xi^tK\left(\mathbb{F}_0(y|\phi)\right) + (1-\eta(y))\xi^tK\left(\hat{\mathbb{F}}_0(y|\phi)\right)\right)\right| \leq M.
\label{eqn:BoundednessPsiConsist}
\end{equation}
This entails using assumption C6 that:
\begin{eqnarray*}
\int{\sqrt{\mathbb{F}_T(y)(1-\mathbb{F}_T(y))}\left|\psi'\left(\eta(y)\xi^tK\left(\mathbb{F}_0(y|\phi)\right) + (1-\eta(y))\xi^tK\left(\hat{\mathbb{F}}_0(y|\phi)\right)\right)\right|dy} & \leq & \\
 M\int{\sqrt{\mathbb{F}_T(y)(1-\mathbb{F}_T(y))}dy} & & \\
 & < & +\infty.
\end{eqnarray*}
Finally, the integral is finite and the compactness of $\Xi$ implies that $\|\xi\|$ is bounded. Therefore, we have:
\[H(\phi,\xi) - H_n(\phi,\xi) = o_P(1),\]
independently of $\xi$ and $\phi$. We may deduce now that:
\[\sup_{\phi,\xi}\left|H(\phi,\xi) - H_n(\phi,\xi)\right| \xrightarrow[n\rightarrow\infty]{\quad \mathbb{P} \quad} 0.\]
This proves assumption A2.\\
Assumption A3 is immediately verified since function $\xi\mapsto H(\phi,\xi)$ is strictly concave. Assumption A4 is what we have assumed in assumption C3. Finally, continuity assumption A5 is a direct result from assumptions C4 and C5 using Lebesgue's continuity theorem. All assumptions of Theorem \ref{theo:MainTheorem} are fulfilled and the consistency of $\hat{\phi}$ follows as a consequence.
\end{proof}
% remark
\begin{remark}
\label{rem:ConsistKdiffChebyshev}
We can prove assumption A2 in the previous proof without the use of the "small o" notation. We first have:
\[\frac{K\left(\hat{\mathbb{F}}_0(y|\phi)\right) - K\left(\mathbb{F}_0(y|\phi)\right)}{\sqrt{\mathbb{F}_T(y)(1-\mathbb{F}_T(y))}} \stackrel{\mathbb{P}}{\rightarrow} 0.\]
This is translated into the following limit:
\[\forall \varepsilon>0, \quad \mathbb{P}\left(\left|\frac{K\left(\hat{\mathbb{F}}_0(y|\phi)\right) - K\left(\mathbb{F}_0(y|\phi)\right)}{\sqrt{\mathbb{F}_T(y)(1-\mathbb{F}_T(y))}}\right|<\varepsilon\right)  \xrightarrow[n\rightarrow\infty]{} 1\]
Thus, there exists a sequence of positive numbers $(a_n)_n$ independent of\footnote{This is possible using Chebyshev's inequality and using the fact that $K\left(\mathbb{F}_0(y|\phi)\right)$ can be bounded independently of $y$ and $\phi$.} $y$ which goes to zero at infinity such that:
\[\mathbb{P}\left(\left|\frac{K\left(\hat{\mathbb{F}}_0(y|\phi)\right) - K\left(\mathbb{F}_0(y|\phi)\right)}{\sqrt{\mathbb{F}_T(y)(1-\mathbb{F}_T(y))}}\right|<\frac{\varepsilon}{\tilde{M}}\right)\geq 1-a_n\]
where $\tilde{M}=M\sup_{\Xi}\|\xi\|\int{\mathbb{F}_T(y)(1-\mathbb{F}_T(y))dy}$ and $M$ is defined through inequality (\ref{eqn:BoundednessPsiConsist}). On the other hand, the event:
\[\left\|\frac{K\left(\hat{\mathbb{F}}_0(y|\phi)\right) - K\left(\mathbb{F}_0(y|\phi)\right)}{\sqrt{\mathbb{F}_T(y)(1-\mathbb{F}_T(y))}}\right\|<\frac{\varepsilon}{\tilde{M}}\]
implies the event:
\begin{eqnarray*}
& & \int{\sqrt{\mathbb{F}_T(y)(1-\mathbb{F}_T(y))}\|\xi\|\left\|\frac{\left(K\left(\mathbb{F}_0(y|\phi)\right) - K\left(\hat{\mathbb{F}}_0(y|\phi)\right)\right)}{\sqrt{\mathbb{F}_T(y)(1-\mathbb{F}_T(y))}}\right\| \psi'\left(\eta(y)\xi^tK\left(\mathbb{F}_0(y|\phi)\right) \right.}\\
&  &  \left.+ (1-\eta(y))\xi^tK\left(\hat{\mathbb{F}}_0(y|\phi)\right)\right)dy \\
&  & <\frac{\varepsilon}{\tilde{M}} \int{\sqrt{\mathbb{F}_T(y)(1-\mathbb{F}_T(y))}\|\xi\| \psi'\left(\eta(y)\xi^tK\left(\mathbb{F}_0(y|\phi)\right) + (1-\eta(y))\xi^tK\left(\hat{\mathbb{F}}_0(y|\phi)\right)\right)dy}\\
& & <  \varepsilon.
\end{eqnarray*}
This entails that 
\[\left|H(\phi,\xi) - H_n(\phi,\xi)\right|<\varepsilon.\]
%\begin{multline*}
%\mathbb{P}\left(\left\|\frac{K\left(\hat{\mathbb{F}}_0(y|\phi)\right) - K\left(\mathbb{F}_0(y|\phi)\right)}{\sqrt{\mathbb{F}_T(y)(1-\mathbb{F}_T(y))}}\right\|<\frac{\varepsilon}{M}\right) \leq \\
%\mathbb{P}\left(\int{\sqrt{\mathbb{F}_T(y)(1-\mathbb{F}_T(y))}\|\xi\|\left\|\frac{\left(K\left(\mathbb{F}_0(y|\phi)\right) - K\left(\hat{\mathbb{F}}_0(y|\phi)\right)\right)}{\sqrt{\mathbb{F}_T(y)(1-\mathbb{F}_T(y))}}\right\| \psi'\left(\eta(y)\xi^tK\left(\mathbb{F}_0(y|\phi)\right) + (1-\eta(y))\xi^tK\left(\hat{\mathbb{F}}_0(y|\phi)\right)\right)dy}\right. < \\
%\left. \frac{\varepsilon}{M} \int{\sqrt{\mathbb{F}_T(y)(1-\mathbb{F}_T(y))}\|\xi\| \psi'\left(\eta(y)\xi^tK\left(\mathbb{F}_0(y|\phi)\right) + (1-\eta(y))\xi^tK\left(\hat{\mathbb{F}}_0(y|\phi)\right)\right)dy}\right) \leq \\
%\mathbb{P}\left(\int{\sqrt{\mathbb{F}_T(y)(1-\mathbb{F}_T(y))}\|\xi\|\left\|\frac{\left(K\left(\mathbb{F}_0(y|\phi)\right) - K\left(\hat{\mathbb{F}}_0(y|\phi)\right)\right)}{\sqrt{\mathbb{F}_T(y)(1-\mathbb{F}_T(y))}}\right\| \psi'\left(\eta(y)\xi^tK\left(\mathbb{F}_0(y|\phi)\right) + (1-\eta(y))\xi^tK\left(\hat{\mathbb{F}}_0(y|\phi)\right)\right)dy}\varepsilon\right) \leq \\
%\mathbb{P}\left(\left|H(\phi,\xi) - H_n(\phi,\xi)\right|<\varepsilon\right).
%\end{multline*}
The final line does not depend on $(\phi,\xi)$, and we may deduce that:
\[\mathbb{P}\left(\sup_{\phi,\xi}\left|H(\phi,\xi) - H_n(\phi,\xi)\right|<\varepsilon\right) \geq 1-a_n.\]
\end{remark}

%%%%%%%%%%%%%%%%%%%%%%%%%%%%%%%%%%%%%%%%%%%%%%%%%%%%%%%%%%%%%%%%%%%%%%%%%%%%%%%%%%%%%%%%%%%%%%%%%%%%%%%
%%%%%%%%%%%%%%%%%%%%%%%%%%%%%%%%%%%%%%%%%%%%%%%%%%%%%%%%%%%%%%%%%%%%%%%%%%%%%%%%%%%%%%%%%%%%%%%%%%%%%%%
\subsection{Proof of Proposition \ref{prop:LimitLawLmomConstrPart}}\label{Append:PropAsyptotNormIntegral}
\begin{proof}
We would like to calculate the difference $\int{K(\hat{\mathbb{F}}_0(y|\phi))dy}-\int{K(\mathbb{F}_0(y|\phi))dy}$ as a functional of the difference $\hat{\mathbb{F}}_0(y|\phi)-\mathbb{F}_0(y|\phi)$. For two reals $a$ and $b$, we have:
\[K_r(a)-K_r(b) = \sum_{k=0}^{r-1}{\frac{c_{r,k}}{k+1}\left(a^{k+1}-b^{k+1}\right)},\]
where $c_{r,k}=(-1)^{r-k-1}\binom{r-1}{k}\binom{r+k-1}{k}$. Using the identity $a^{k+1}-b^{k+1} = (a-b)\sum_{j=0}^{k}{a^jb^{k-j}}$, we can write:
\[K_r(a)-K_r(b) = (a-b)\sum_{k=0}^{r-1}\sum_{j=0}^{k}{\frac{c_{r,k}}{k+1}a^jb^{k-j}}.\]
Applying this formula on $a=\hat{\mathbb{F}}_0(y|\phi)$ and $b=\mathbb{F}_0(y|\phi)$ yields
\begin{eqnarray}
K_r\left(\hat{\mathbb{F}}_0(y|\phi)\right)-K_r\left(\mathbb{F}_0(y|\phi)\right) & = & \left(\hat{\mathbb{F}}_0(y|\phi)-\mathbb{F}_0(y|\phi)\right)\sum_{k=0}^{r-1}\sum_{j=0}^{k}{\frac{c_{r,k}}{k+1}\hat{\mathbb{F}}_0(y|\phi)^j \mathbb{F}_0(y|\phi)^{k-j}} \nonumber \\
& = & \frac{1}{1-\lambda}\left(\mathbb{F}_n(y)-\mathbb{F}_T(y)\right)\sum_{k=0}^{r-1}\sum_{j=0}^{k}{\frac{c_{r,k}}{k+1}\hat{\mathbb{F}}_0(y|\phi)^j \mathbb{F}_0(y|\phi)^{k-j}}. \nonumber \\
\label{eqn:Kdifference}
\end{eqnarray}
We will show that the sum term can be rewritten using only $\mathbb{F}_0(y|\phi)$. By the Kolmogorov-Smirnov theorem, we have:
\[\sup_{y}\left|\hat{\mathbb{F}}_0(y|\phi)-\mathbb{F}_0(y|\phi)\right| = \sup_{y}\left|\mathbb{F}_n(y)-\mathbb{F}_T(y)\right|  = O_P\left(\frac{1}{\sqrt{n}}\right).\]
This permits us to simply write that
\[\hat{\mathbb{F}}_0(y|\phi)=\mathbb{F}_0(y|\phi) + O_P\left(\frac{1}{\sqrt{n}}\right),\]
with $O_P\left(\frac{1}{\sqrt{n}}\right)$ tends to zero in probability as $n$ goes to infinity independently of $y$. Thus formula (\ref{eqn:Kdifference}) can be rewritten as:
\begin{eqnarray*}
K_r\left(\hat{\mathbb{F}}_0(y|\phi)\right)-K_r\left(\mathbb{F}_0(y|\phi)\right) & = & \frac{1}{1-\lambda}\left(\mathbb{F}_n(y)-\mathbb{F}_T(y)\right)\sum_{k=0}^{r-1}\sum_{j=0}^{k}{\frac{c_{r,k}}{k+1}\left(\mathbb{F}_0(y|\phi)^j + O_P\left(\frac{1}{\sqrt{n}}\right)\right) }\\ 
 & & \text{\vspace{3cm}} \times \mathbb{F}_0(y|\phi)^{k-j}\\
 & = & \frac{1}{1-\lambda}\left(\mathbb{F}_n(y)-\mathbb{F}_T(y)\right)\sum_{k=0}^{r-1}{c_{r,k}\mathbb{F}_0(y|\phi)^k} \\
 & &   + O_P\left(\frac{1}{\sqrt{n}}\right) \frac{1}{1-\lambda}\left(\mathbb{F}_n(y)-\mathbb{F}_T(y)\right)\sum_{k=0}^{r-1}\sum_{j=0}^{k}{\frac{c_{r,k}}{k+1} \mathbb{F}_0(y|\phi)^{k-j}}.
\end{eqnarray*}
Integrating the two sides of the previous equation and multiplying by $\sqrt{n}$ gives:
\begin{multline}
\sqrt{n}\int{\left[K_r\left(\hat{\mathbb{F}}_0(y|\phi)\right)-K_r\left(\mathbb{F}_0(y|\phi)\right)\right]dy} = \frac{1}{1-\lambda}\int{\sqrt{n}\left(\mathbb{F}_n(y)-\mathbb{F}_T(y)\right)\sum_{k=0}^{r-1}{c_{r,k}\mathbb{F}_0(y|\phi)^k}dy}\\ + O_P\left(\frac{1}{\sqrt{n}}\right)\frac{1}{1-\lambda}\int{\sqrt{n}\left(\mathbb{F}_n(y)-\mathbb{F}_T(y)\right)\sum_{k=0}^{r-1}\sum_{j=0}^{k}{\frac{c_{r,k}}{k+1} \mathbb{F}_0(y|\phi)^{k-j}} dy}.
\label{eqn:IntegDiffK}
\end{multline}
The first integral in the right hand side is the part which will produce the Gaussian distribution of the limit law using the CLT. It remains to prove that the second integral in the right hand side tends to zero in probability. Using the law of iterated logarithm, we can write:
\begin{equation}
\limsup_{n\rightarrow\infty}\sqrt{\frac{n}{\log\log n}}\frac{\mathbb{F}_n(y)-\mathbb{F}_T(y)}{\sqrt{\mathbb{F}_T(y)\left(1-\mathbb{F}_T(y)\right)}} = \sqrt{2}.
\label{eqn:IterLogLaw}
\end{equation}
We now may write the integral in the second term as follows:
\begin{multline*}
O_P\left(\frac{1}{\sqrt{n}}\right)\int{\sqrt{n}\left(\mathbb{F}_n(y)-\mathbb{F}_T(y)\right)\sum_{k=0}^{r-1}\sum_{j=0}^{k}{\frac{c_{r,k}}{k+1} \mathbb{F}_0(y|\phi)^{k-j}} dy} = \\ 
O_P\left(\sqrt{\frac{\log\log n}{n}}\right)\int{\sqrt{\frac{n}{\log\log n}}\frac{\mathbb{F}_n(y)-\mathbb{F}_T(y)}{\sqrt{\mathbb{F}_T(y)\left(1-\mathbb{F}_T(y)\right)}} \sqrt{\mathbb{F}_T(y)\left(1-\mathbb{F}_T(y)\right)}  \sum_{k=0}^{r-1}\sum_{j=0}^{k}{\frac{c_{r,k}}{k+1} \mathbb{F}_0(y|\phi)^{k-j}} dy}.
\end{multline*}
The sum term inside the integral is bounded uniformly on $y$. Combine this with the limit in (\ref{eqn:IterLogLaw}), we may deduce that for $n$ sufficiently large, there exists a constant $M$ such that:
\begin{eqnarray*}
\int{\sqrt{\frac{n}{\log\log n}}\frac{\left|\mathbb{F}_n(y)-\mathbb{F}_T(y)\right|}{\sqrt{\mathbb{F}_T(y)\left(1-\mathbb{F}_T(y)\right)}} \sqrt{\mathbb{F}_T(y)\left(1-\mathbb{F}_T(y)\right)}  \sum_{k=0}^{r-1}\sum_{j=0}^{k}{\frac{|c_{r,k}|}{k+1} \mathbb{F}_0(y|\phi)^{k-j}} dy} & \leq & \\
 M\int{\sqrt{\mathbb{F}_T(y)\left(1-\mathbb{F}_T(y)\right)} dy} & &  \\
& < & \infty
\end{eqnarray*}
Thus, the integral exists and is finite for sufficiently large $n$. This entails that:
\begin{equation}
O_P\left(\frac{1}{\sqrt{n}}\right)\int{\sqrt{n}\left(\mathbb{F}_n(y)-\mathbb{F}_T(y)\right)\sum_{k=0}^{r-1}\sum_{j=0}^{k}{\frac{c_{r,k}}{k+1} \mathbb{F}_0(y|\phi)^{k-j}} dy} \xrightarrow[\mathbb{P}]{\quad n\rightarrow\infty\quad} 0,\quad \text{in probability.}
\label{eqn:LimitLawPart2}
\end{equation}
Going back to equation (\ref{eqn:IntegDiffK}), the second term in the right hand side tends to zero in probability. We need now to treat the first term.
\begin{equation}
\int{\sqrt{n}\left(\mathbb{F}_n(y)-\mathbb{F}_T(y)\right)\sum_{k=0}^{r-1}{c_{r,k}\mathbb{F}_0(y|\phi)^k}} = \frac{1}{\sqrt{n}}\sum_{i=1}^n{\int{\left(\ind{X_i\leq y} - \mathbb{F}_T(y)\right)}\sum_{k=0}^{r-1}{c_{r,k}\mathbb{F}_0(y|\phi)^k}dy}.
\label{eqn:AsymptotNormGaussSumCLT}
\end{equation}
This is a sum of i.i.d. random variables. Before proceeding any further, it is necessary to prove that such random variables are well defined (the integrals exist) and have a finite variance. First of all, we have:
\begin{multline}
\int_{-\infty}^{\infty}{\left|\ind{X_i\leq y} - \mathbb{F}_T(y)\right|\sum_{k=0}^{r-1}{c_{r,k}\mathbb{F}_0(y|\phi)^k}dy} =  \int_{-\infty}^{X_i}{\mathbb{F}_T(y)}\sum_{k=0}^{r-1}{c_{r,k}\mathbb{F}_0(y|\phi)^kdy} \\ + \int_{X_i}^{\infty}{\left(1 - \mathbb{F}_T(y)\right)\sum_{k=0}^{r-1}{c_{r,k}\mathbb{F}_0(y|\phi)^k}dy}
\label{eqn:AsymptoNormLmomIntegrability}
\end{multline}
On the other hand, since the $|X_i|$'s have finite expectation, then $X_i$ is finite almost surely and we have:
\begin{eqnarray*}
\mathbb{E}|X_i| & = & \int_{t=0}^{\infty}{\mathbb{P}\left(|X_i|>t\right)dt} \\
 & = & \int_{0}^{\infty}{\left(1-\mathbb{F}_T(t)\right)dt} + \int_{-\infty}^{0}{\mathbb{F}_T(t)dt}.
\end{eqnarray*}
Thus, $\mathbb{F}_T(t)$ is integrable in the neighborhood of $-\infty$, and $1-\mathbb{F}_T(t)$ is integrable in the neighborhood of $+\infty$. This proves that the integral in equation (\ref{eqn:AsymptoNormLmomIntegrability}) exists and is finite. Now the random variables in (\ref{eqn:AsymptotNormGaussSumCLT}) are well defined. The expectation is zero using the Fubini's theorem:
\[\mathbb{E}\left[\int{\left(\ind{X_i\leq y} - \mathbb{F}_T(y)\right)\sum_{k=0}^{r-1}{c_{r,k}\mathbb{F}_0(y|\phi)^k}dy}\right] = \int{\mathbb{E}\left(\ind{X_i\leq y} - \mathbb{F}_T(y)\right)\sum_{k=0}^{r-1}{c_{r,k}\mathbb{F}_0(y|\phi)^k}dy}=0.\]
The final part of the proof is to calculate the covariance matrix. Let $r_1$ and $r_2$ be two positive natural numbers such that $r_1\leq\ell$ and $r_2\leq\ell$. The Fubini's theorem yields:
\begin{multline*}
\mathbb{E}\int{\left(\ind{X_i\leq y} - \mathbb{F}_T(y)\right)\sum_{k=0}^{r_1-1}{c_{r_1,k}\mathbb{F}_0(y|\phi)^k}dy}\int{\left(\ind{X_i\leq x} - \mathbb{F}_T(x)\right)\sum_{k=0}^{r_2-1}{c_{r_2,k}\mathbb{F}_0(x|\phi)^k}dx} = \\
\int{\int{\mathbb{E}\left(\ind{X_i\leq x} - \mathbb{F}_T(x)\right)\left(\ind{X_i\leq y} - \mathbb{F}_T(y)\right)\sum_{k=0}^{r_1-1}{c_{r_1,k}\mathbb{F}_0(x|\phi)^k}\sum_{k=0}^{r_2-1}{c_{r_2,k}\mathbb{F}_0(y|\phi)^k}dy}dx}
\end{multline*}
Denoting $\Sigma$ the covariance matrix, we may write:
\[\Sigma_{r_1,r_2} = \int{\int{\left(\mathbb{F}_T\left(\min(x,y)\right) - \mathbb{F}_T(x)\mathbb{F}_T(y)\right)\sum_{k=0}^{r_1-1}{c_{r_1,k}\mathbb{F}_0(x|\phi)^k}\sum_{k=0}^{r_2-1}{c_{r_2,k}\mathbb{F}_0(y|\phi)^k}dy}dx}.\]
The sum of i.i.d. variables in (\ref{eqn:AsymptotNormGaussSumCLT}) are now well defined and the CLT applies and gives:
\[\frac{1}{\sqrt{n}}\sum_{i=1}^n{\int{\left(\ind{X_i\leq y} - \mathbb{F}_T(y)\right)}\sum_{k=0}^{r-1}{c_{r,k}\mathbb{F}_0(y|\phi)^k}dy} \xrightarrow[]{\quad \mathcal{D}\quad} \mathcal{N}(0,\Sigma).\]
This result together with (\ref{eqn:LimitLawPart2}) and (\ref{eqn:IntegDiffK}) complete the proof.
\end{proof}

%%%%%%%%%%%%%%%%%%%%%%%%%%%%%%%%%%%%%%%%%%%%%%%%%%%%%%%%%%%%%%%%%%%%%%%%%%%%%%%%%%%%%%%%%%%%%%%%%%%%%%%
%%%%%%%%%%%%%%%%%%%%%%%%%%%%%%%%%%%%%%%%%%%%%%%%%%%%%%%%%%%%%%%%%%%%%%%%%%%%%%%%%%%%%%%%%%%%%%%%%%%%%%%
\subsection{Proof of Theorem \ref{theo:AsymptotNormLmom}}\label{Append:TheoNormalAsymptotLmom}
\begin{proof}
The proof is based on a mean value expansion between $(\hat{\phi},\xi_n(\hat{\phi}))$ and $(\phi^*,0)$. We therefore, need to calculate the first and second order derivatives.\\ 
First order derivatives are given by:
\begin{eqnarray*}
\frac{\partial H_n}{\partial \xi}(\phi,\xi) & = & m(\alpha) - \int{K(\hat{\mathbb{F}}_0(y|\phi)) \psi'\left(\xi^tK(\hat{\mathbb{F}}_0(y|\phi))\right)dy} \\
\frac{\partial H_n}{\partial \alpha}(\phi,\xi) & = & \xi^t\nabla m(\alpha) \\
\frac{\partial H_n}{\partial \lambda}(\phi,\xi) & = & -\int{\left[\frac{1}{(1-\lambda)^2}\mathbb{F}_n(y) - \frac{1}{(1-\lambda)^2}\mathbb{F}_1(y|\theta)\right]\xi^tK'(\hat{\mathbb{F}}_0(y|\phi))\psi'\left(\xi^t K(\hat{\mathbb{F}}_0(y|\phi))\right)dy} \\
\frac{\partial H_n}{\partial \theta}(\phi,\xi) & = & \frac{\lambda}{1-\lambda} \int{\nabla_{\theta}\mathbb{F}_1(y|\theta)\xi^t K'(\hat{\mathbb{F}}_0(y|\phi)) \psi'\left(\xi^t K(\hat{\mathbb{F}}_0(y|\phi))\right)dy}.
\end{eqnarray*}
Second order derivatives are given by:
\begin{eqnarray*}
\frac{\partial^2 H_n}{\partial \xi^2}(\phi,\xi) & = & \int{K(\hat{\mathbb{F}}_0(y|\phi))K(\hat{\mathbb{F}}_0(y|\phi))^t \psi''\left(\xi^tK(\hat{\mathbb{F}}_0(y|\phi))\right)dy} \\
\frac{\partial^2 H_n}{\partial \alpha^2}(\phi,\xi) & = & \xi^t J_{m(\alpha)} \\
\frac{\partial^2 H_n}{\partial^2 \lambda}(\phi,\xi) & = & -\int{\left[\frac{2}{(1-\lambda)^3}\mathbb{F}_n(y) - \frac{2}{(1-\lambda)^3}\mathbb{F}_1(y|\theta)\right]\xi^tK'(\hat{\mathbb{F}}_0(y|\phi))\psi'\left(\xi^t K(\hat{\mathbb{F}}_0(y|\phi))\right)dy} \\  
					& & -\int{\left[\frac{1}{(1-\lambda)^2}\mathbb{F}_n(y) - \frac{1}{(1-\lambda)^2}\mathbb{F}_1(y|\theta)\right]^2\xi^tK''(\hat{\mathbb{F}}_0(y|\phi))\psi'\left(\xi^t K(\hat{\mathbb{F}}_0(y|\phi))\right)dy} \\
					&  &	-\int{\left[\frac{1}{(1-\lambda)^2}\mathbb{F}_n(y) - \frac{1}{(1-\lambda)^2}\mathbb{F}_1(y|\theta)\right]^2\left[(\xi^tK'(\hat{\mathbb{F}}_0(y|\phi))\right]^2\psi''\left(\xi^t K(\hat{\mathbb{F}}_0(y|\phi))\right)dy} \\
\frac{\partial^2 H_n}{\partial \theta^2}(\phi,\xi) & = & \frac{\lambda}{1-\lambda} \int{J_{\mathbb{F}_1(.|\theta)}\xi^t K'(\hat{\mathbb{F}}_0(y|\phi)) \psi'\left(\xi^t K(\hat{\mathbb{F}}_0(y|\phi))\right)dy} \\
				&  & - \frac{\lambda^2}{(1-\lambda)^2} \int{\nabla_{\theta}\mathbb{F}_1(y|\theta)\nabla_{\theta}\mathbb{F}_1(y|\theta)^t \xi^t K''(\hat{\mathbb{F}}_0(y|\phi)) \psi'\left(\xi^t K(\hat{\mathbb{F}}_0(y|\phi))\right)dy} \\
				 &  & -\frac{\lambda^2}{(1-\lambda)^2} \int{\nabla_{\theta}\mathbb{F}_1(y|\theta)\nabla_{\theta}\mathbb{F}_1(y|\theta)^t \left[\xi^t K'(\hat{\mathbb{F}}_0(y|\phi))\right]^2 \psi''\left(\xi^t K(\hat{\mathbb{F}}_0(y|\phi))\right)dy}
\end{eqnarray*}
Crossed derivatives:
\begin{eqnarray*}
\frac{\partial^2 H_n}{\partial \xi \partial\alpha}(\phi,\xi) & = & \nabla m(\alpha) \\
\frac{\partial^2 H_n}{\partial \xi \partial\lambda}(\phi,\xi) & = &  -\int{\left[\frac{1}{(1-\lambda)^2}\mathbb{F}_n(y) - \frac{1}{(1-\lambda)^2}\mathbb{F}_1(y|\theta)\right]K'(\hat{\mathbb{F}}_0(y|\phi))\psi'\left(\xi^t K(\hat{\mathbb{F}}_0(y|\phi))\right)dy} - \\
 &  & \int{K(\hat{\mathbb{F}}_0(y|\phi))\left[\frac{1}{(1-\lambda)^2}\mathbb{F}_n(y) - \frac{1}{(1-\lambda)^2}\mathbb{F}_1(y|\theta)\right]\xi^tK'(\hat{\mathbb{F}}_0(y|\phi))\psi'\left(\xi^t K(\hat{\mathbb{F}}_0(y|\phi))\right)dy} \\
\frac{\partial^2 H_n}{\partial \xi \partial\theta}(\phi,\xi) & = & \frac{\lambda}{1-\lambda} \int{\nabla_{\theta}\mathbb{F}_1(y|\theta)K'(\hat{\mathbb{F}}_0(y|\phi))^t \psi'\left(\xi^t K(\hat{\mathbb{F}}_0(y|\phi))\right)dy} \\
 &  & + \frac{\lambda}{1-\lambda} \int{K(\hat{\mathbb{F}}_0(y|\phi))\nabla_{\theta}\mathbb{F}_1(y|\theta)^t\xi^t K'(\hat{\mathbb{F}}_0(y|\phi)) \psi'\left(\xi^t K(\hat{\mathbb{F}}_0(y|\phi))\right)dy}\\
\frac{\partial^2 H_n}{\partial \alpha \partial\lambda}(\phi,\xi) & = & 0 \\
\frac{\partial^2 H_n}{\partial \alpha \partial\theta}(\phi,\xi) & = & 0 \\
\frac{\partial^2 H_n}{\lambda \partial\theta}(\phi,\xi) & = & \frac{1}{(1-\lambda)^2}\int{\nabla\mathbb{F}_1(y|\theta)\xi^tK'(\hat{\mathbb{F}}_0(y|\phi))\psi'\left(\xi^t K(\hat{\mathbb{F}}_0(y|\phi))\right)dy} \\
	&  &  + \frac{\lambda}{1-\lambda} \int{\nabla_{\theta}\mathbb{F}_1(y|\theta) \left[\frac{1}{(1-\lambda)^2}\mathbb{F}_n(y) - \frac{1}{(1-\lambda)^2}\mathbb{F}_1(y|\theta)\right]\xi^t K''(\hat{\mathbb{F}}_0(y|\phi))} \\
	& & \qquad \times \psi'\left(\xi^t K(\hat{\mathbb{F}}_0(y|\phi))\right)dy \\  
	&  & + \frac{\lambda}{1-\lambda} \int{\nabla_{\theta}\mathbb{F}_1(y|\theta) \left[\frac{1}{(1-\lambda)^2}\mathbb{F}_n(y) - \frac{1}{(1-\lambda)^2}\mathbb{F}_1(y|\theta)\right] \left[\xi^t K'(\hat{\mathbb{F}}_0(y|\phi))\right]^2} \\
	 & & \qquad \times \psi''\left(\xi^t K(\hat{\mathbb{F}}_0(y|\phi))\right)dy
\end{eqnarray*}
Notice that by assumption 1, interesting values of $\xi$ are only in a neighborhood of the vector 0 which can be taken to be the ball $B(0,\varepsilon)$ for some $\varepsilon>0$. Besides, the derivatives given here above are well defined using Lebesgue theorems. Indeed, all integrands are controlled by either $K(\hat{\mathbb{F}}(y|\phi))$ or $\mathbb{F}_n(y)-\mathbb{F}_1(.|\theta)$ which are both integrable independently of $\phi$ as soon as $\mathbb{F}_1(.|\theta)$ has a finite expectation. Similar discussion for the former was given in Example \ref{Example:Chi2Lmom}, and for the later in the proof of Proposition \ref{prop:LimitLawLmomConstrPart} but for $\mathbb{F}_n(y)-\mathbb{F}_T(.|\theta)$ instead. Other derivatives are controlled by assumtions 4-6 of the present theorem.\\
A mean value expansion of the gradient of $H_n$ between $(\hat{\phi},\xi_n(\hat{\phi}))$ with Lagrange remainder gives that there exists $(\bar{\phi},\bar{\xi})$ on the line between these two points such that:
\begin{equation}
\left(\begin{array}{c} \frac{\partial H_n}{\partial \phi}(\hat{\phi},\xi(\hat{\phi})) \\ \frac{\partial H_n}{\partial \xi}(\hat{\phi},\xi_n(\hat{\phi})) \end{array}\right) = \left(\begin{array}{c}  \frac{\partial H_n}{\partial \phi}(\phi^*,0) \\\frac{\partial H_n}{\partial \xi}(\phi^*,0) \end{array}\right)  + J_{H_n}(\bar{\phi},\bar{\xi}) \left(\begin{array}{c} \hat{\phi}-\phi^* \\ \xi_n(\hat{\phi})\end{array}\right),
\label{eqn:StochExpansionLmom}
\end{equation}
where $J_{H_n}(\bar{\phi},\bar{\xi})$ is the matrix of second derivatives of $H_n$ calculated at the mid point $(\bar{\phi},\bar{\xi})$. First order optimality condition at $(\hat{\phi},\xi_n(\hat{\phi}))$ is translated by:
\begin{eqnarray*}
\frac{\partial}{\partial \xi} H_n(\hat{\phi},\xi_n(\hat{\phi})) & = & 0 \\
\left.\frac{\partial}{\partial \phi}\left(H_n(\phi,\xi_n(\phi))\right)\right|_{\phi=\hat{\phi}} & = & 0. 
\end{eqnarray*}
The chain rule permits us to calculate the second line simply as a derivative with respect to $\phi$ calculated at the optimal point $(\hat{\phi},\xi_n(\hat{\phi}))$, i.e.
\begin{eqnarray*}
\left.\frac{\partial}{\partial \phi}\left(H_n(\phi,\xi_n(\phi))\right)\right|_{\phi=\hat{\phi}} & = &  \frac{\partial}{\partial \phi}H_n(\hat{\phi},\xi_n(\hat{\phi})) + \frac{\partial}{\partial \xi} H_n(\hat{\phi},\xi_n(\hat{\phi})) \frac{\partial \xi_n}{\partial \phi}(\hat{\phi}) \\
 & = & \frac{\partial}{\partial \phi}H_n(\hat{\phi},\xi_n(\hat{\phi})).
\end{eqnarray*}
Thus, optimality conditions at $(\hat{\phi},\xi_n(\hat{\phi}))$ are given by:
\[\frac{\partial H_n}{\partial \xi}(\hat{\phi},\xi_n(\hat{\phi})) = 0, \quad 
\frac{\partial H_n}{\partial \alpha}(\hat{\phi},\xi_n(\hat{\phi})) = 0, \quad
\frac{\partial H_n}{\partial \lambda}(\hat{\phi},\xi_n(\hat{\phi})) = 0, \quad
\frac{\partial H_n}{\partial \theta}(\hat{\phi},\xi_n(\hat{\phi})) = 0.
\]
On the other hand, we have at $(\phi^*,0)$:
\begin{eqnarray*}
\frac{\partial H_n}{\partial \xi}(\phi^*,0) = m(\alpha^*) - \int{K(\hat{\mathbb{F}}_0(y|\phi^*))dy},
\frac{\partial H_n}{\partial \alpha}(\phi^*,0) = 0,
\frac{\partial H_n}{\partial \lambda}(\phi^*,0) = 0,
\frac{\partial H_n}{\partial \theta}(\phi^*,0) =  0.
\end{eqnarray*}
By proposition \ref{prop:LimitLawLmomConstrPart}, since $m(\alpha^*) = \int{K(\mathbb{F}_0(y|\phi^*))}$,
\begin{equation}
\sqrt{n}\left[m(\alpha^*) - \int{K(\hat{\mathbb{F}}_0(y|\phi^*))dy}\right] \xrightarrow{\mathcal{L}}{} \mathcal{N}(0,\Sigma)
\label{eqn:LimitLawPartialDerivHnLmom}
\end{equation}
with $\Sigma$ is the matrix of covariance defined by formula (\ref{eqn:VarCovMatConstrPart}). It remains now to calculate the limit in probability of the matrix $J_{H_n}(\bar{\phi},\bar{\xi})$. Recall first that as $n$ goes to infinity $\bar{\phi}\rightarrow\phi^*$ and $\bar{\xi}\rightarrow 0$. Moreover, by the Slutsky theorem and the law of large numbers, we have:
\[\hat{\mathbb{F}}_0(y,|\bar{\phi}) = \frac{1}{1-\bar{\lambda}} \mathbb{F}_n(y) - \frac{\bar{\lambda}}{1-\bar{\lambda}} \mathbb{F}_1(y|\bar{\theta})\xrightarrow{n\rightarrow\infty}{} \frac{1}{1-\lambda^*} \mathbb{F}_T(y) - \frac{\lambda^*}{1-\lambda^*} \mathbb{F}_1(y|\theta^*) = \mathbb{F}_0(y|\phi^*).\]
We may now give the limit of the blocs of the matrix $J_{H_n}(\bar{\phi},\bar{\xi})$:
\begin{eqnarray*}
\frac{\partial^2 H_n}{\partial \xi^2}(\phi^*,0) =  \int{K(\mathbb{F}_0(y|\phi^*))K(\mathbb{F}_0(y|\phi^*))^tdy}, \qquad
\frac{\partial^2 H_n}{\partial \alpha^2}(\phi^*,0) = 0, \\
\frac{\partial^2 H_n}{\partial^2 \lambda}(\phi^*,0)  =  0, \qquad 
\frac{\partial^2 H_n}{\partial \theta^2}(\phi^*,0) = 0.
\end{eqnarray*}
Crossed derivatives:
\[\frac{\partial^2 H_n}{\partial \xi \partial\alpha}(\phi^*,0) = \nabla m(\alpha^*),\quad \frac{\partial^2 H_n}{\partial \alpha \partial\lambda}(\phi^*,0) = 0,\quad \frac{\partial^2 H_n}{\partial \alpha \partial\theta}(\phi^*,0) = 0  ,\quad 
\frac{\partial^2 H_n}{\lambda \partial\theta}(\phi^*,0) = 0, \]
\[\frac{\partial^2 H_n}{\partial \xi \partial\lambda}(\phi^*,0) = -\int{\left[\frac{1}{(1-\lambda^*)^2}\mathbb{F}_T(y) - \frac{1}{(1-\lambda^*)^2}\mathbb{F}_1(y|\theta^*)\right]K'(\mathbb{F}_0(y|\phi^*))dy}\]
\[\frac{\partial^2 H_n}{\partial \xi \partial\theta}(\phi^*,0) = \frac{\lambda^*}{1-\lambda^*} \int{\nabla_{\theta}\mathbb{F}_1(y|\theta^*)K'(\mathbb{F}_0(y|\phi^*))^tdy}.\]
The limit in probability of the matrix $J_{H_n}(\bar{\phi},\bar{\xi})$ can be written in the form:
\[ J_H = \left[\begin{array}{cc}
0 & J_{\phi^*,\xi^*}^t \\
J_{\phi^*,\xi^*} & J_{\xi^*,\xi^*}
\end{array}\right]\]
where $J_{\phi^*,\xi^*}$ and $J_{\xi^*,\xi^*}$ are given by (\ref{eqn:NormalAsymLMomJ1}) and (\ref{eqn:NormalAsymLMomJ2}). The inverse of matrix $J_H$ has the form:
\[J_H^{-1} = \left(\begin{array}{cc} -\tilde{\Sigma} & H \\ H^t & P\end{array}\right),\]
where
\[
\tilde{\Sigma} = \left(J_{\phi^*,\xi^*}^t J_{\xi^*,\xi^*} J_{\phi^*,\xi^*}\right)^{-1},\quad  H = \tilde{\Sigma} J_{\phi^*,\xi^*}^t J_{\xi^*,\xi^*}^{-1},\quad  P = J_{\xi^*,\xi^*}^{-1} - J_{\xi^*,\xi^*}^{-1} J_{\phi^*,\xi^*} \tilde{\Sigma} J_{\phi^*,\xi^*}^t J_{\xi^*,\xi^*}^{-1}
\]
Going back to (\ref{eqn:StochExpansionLmom}), we have:
\begin{equation*}
\left(\begin{array}{c} 0 \\ 0 \end{array}\right) = \left(\begin{array}{c}  0 \\ \frac{\partial H_n}{\partial \xi}(\phi^*,0) \end{array}\right)  + J_{H_n}(\bar{\phi},\bar{\xi}) \left(\begin{array}{c}  \hat{\phi}-\phi^* \\ \xi_n(\hat{\phi}) \end{array}\right).
\end{equation*}
Solving this equation in $\phi$ and $\xi$ gives:
\begin{equation*}
\left(\begin{array}{c}  \sqrt{n}\left(\hat{\phi}-\phi^*\right) \\ \sqrt{n}\xi_n(\hat{\phi})\end{array}\right) = J_H^{-1}\left(\begin{array}{c}  0 \\ \sqrt{n}\frac{\partial H_n}{\partial \xi}(\phi^*,0) \end{array}\right) + o_P(1).
\end{equation*}
Finally, using (\ref{eqn:LimitLawPartialDerivHnLmom}), we get that:
\[\left(\begin{array}{c}  \sqrt{n}\left(\hat{\phi}-\phi^*\right) \\ \sqrt{n}\xi_n(\hat{\phi})\end{array}\right) \xrightarrow[\mathcal{L}]{} \mathcal{N}\left(0,S\right)\]
where 
\[S=\left(\begin{array}{c}H \\ P\end{array}\right) \Sigma \left(H^t\quad P^t\right).\]
This ends the proof.
\end{proof}

\bibliographystyle{plainnat}
\bibliography{bibliography-paper}

\end{document}